\documentclass[twocolumn,secnumarabic,amssymb, nobibnotes, prl,superscriptaddress]{revtex4-2} 
\usepackage{CJK}
\usepackage{graphicx} 
\usepackage{blindtext}
\usepackage{physics} 

\usepackage{lipsum}
\usepackage{amsfonts,amssymb,amsmath}
\usepackage{bm}
\usepackage{xcolor,calc}
\usepackage{ulem}
\usepackage{hyperref}
\usepackage{siunitx}

\newcommand{\gev}{GeV }
\newcommand{\siv}{SiV }

\newcommand{\nuc}{${}^{13}$C }
\newcommand{\nucc}{${}^{13}$C}

\begin{document}

\title{Coherent Control of a Long-Lived Nuclear Memory Spin in a Germanium-Vacancy Multi-Qubit Node} 

\author{Nick Grimm}
\thanks{These authors contributed equally to this work. \\Corresponding author: \href{mailto:katharina.senkalla@uni-ulm.de}{katharina.senkalla@uni-ulm.de}}
\affiliation{Institute for Quantum Optics, Ulm University, Albert-Einstein-Allee 11, D-89081 Ulm, Germany}
\author{Katharina Senkalla}
\thanks{These authors contributed equally to this work. \\Corresponding author: \href{mailto:katharina.senkalla@uni-ulm.de}{katharina.senkalla@uni-ulm.de}}
\affiliation{Institute for Quantum Optics, Ulm University, Albert-Einstein-Allee 11, D-89081 Ulm, Germany}
\author{Philipp J. Vetter}
\thanks{These authors contributed equally to this work. \\Corresponding author: \href{mailto:katharina.senkalla@uni-ulm.de}{katharina.senkalla@uni-ulm.de}}
\affiliation{Institute for Quantum Optics, Ulm University, Albert-Einstein-Allee 11, D-89081 Ulm, Germany}
\author{Jurek Frey}
\affiliation{Peter Gr\"unberg Institute-Quantum Computing Analytics (PGI-12), Forschungszentrum J\"ulich GmbH, D-52425 J\"ulich, Germany}
\affiliation{Theoretical Physics, Saarland University, D-66123 Saarbrücken, Germany}
\author{Prithvi Gundlapalli}
\affiliation{Institute for Quantum Optics, Ulm University, Albert-Einstein-Allee 11, D-89081 Ulm, Germany}
\author{Tommaso Calarco}
\affiliation{Peter Gr\"unberg Institute-Quantum Control (PGI-8), Forschungszentrum J\"ulich GmbH, D-52425 J\"ulich, Germany}
\affiliation{Institute for Theoretical Physics, University of Cologne, D-50937 Cologne, Germany}
\affiliation{Dipartimento di Fisica e Astronomia, Universit\`a di Bologna, 40127 Bologna, Italy}
\author{Genko Genov}
\affiliation{Institute for Quantum Optics, Ulm University, Albert-Einstein-Allee 11, D-89081 Ulm, Germany}
\author{Matthias M. M\"uller}
\affiliation{Peter Gr\"unberg Institute-Quantum Control (PGI-8), Forschungszentrum J\"ulich GmbH, D-52425 J\"ulich, Germany}
\author{Fedor Jelezko}
\affiliation{Institute for Quantum Optics, Ulm University, Albert-Einstein-Allee 11, D-89081 Ulm, Germany}

\date{\today}
    
\begin{abstract}
The ability to process and store information on surrounding nuclear spins is a major requirement for group-IV color center-based repeater nodes.
We demonstrate coherent control of a \nuc nuclear spin strongly coupled to a negatively charged germanium-vacancy center in diamond with coherence times beyond 2.5\,s at mK temperatures, which is the longest reported for group-IV defects.
Detailed analysis allows us to model the system's dynamics, extract the coupling parameters, and characterize noise.
We estimate an achievable memory time of 18.1\,s with heating limitations considered, paving the way to successful applications as a quantum repeater node.
\end{abstract}

\maketitle


Long-distance quantum communication paves the way to advanced quantum technologies including blind and distributed quantum computing for processing quantum information \cite{ nickerson2014freely, choi2019percolation,cuomo2020towards,drmota2024verifiable}, quantum key distribution for secure communication \cite{ekert1991quantum,elliott2002building,lo2014secure, xu2020secure}, and quantum-enhanced metrology \cite{gottesman2012longer, komar2014quantum, khabiboulline2019optical, yang2024quantum}.
A key element to achieve this is the quantum repeater to mitigate losses in quantum channels~\cite{briegel1998quantum}. \\
Group-IV color centers in diamond have emerged as promising candidates for such repeater nodes \cite{ruf2021quantum} due to their efficient spin-photon interface~\cite{rogers2014alloptical, siyushev2017optical, gorlitz2020spectroscopic,wang2024transform}.
This interface can be further enhanced by integration into nanophotonic devices and cavities~\cite{bradac2019quantum,sipahigil2016integrated,bhaskar2017quantum, rugar2021quantum, antoniuk2024all, parker2024diamond,riegel24lifetimereduction}.
To generate entanglement between distant nodes, these systems require long coherence times that are primarily limited by phonon-mediated dephasing in group-IV defects~\cite{pingault2017coherent,harris2024coherence}.
Innovations such as strain engineering~\cite{meesala18strainengineering,sohn2018controlling,stas2022robust} and the use of dilution refrigerators~\cite{zu2022development} have enabled coherence times of up to 20\,ms \cite{sukachev2017silicon,senkalla2024germanium,karapatzakis2024microwavecontroltinvacancyspin}.
Another key requirement is the ability to address nuclear spins, which can serve as long-lived quantum memory and thus extend the entanglement distance.
This has been effectively demonstrated with the silicon-vacancy center (SiV), which achieved entanglement across a metropolitan fiber network by utilizing the inherent $^{29}\text{Si}$ nuclear spin \cite{knaut2024entanglement}.\\
On the other hand surrounding \nuc nuclear spins can also act as memory qubits across all group-IV platforms and additionally can be used to form a register.
Such quantum register would allow for quantum error correction (QEC)~\cite{taminiau2014universal, unden2016quantum} and entanglement purification protocols~\cite{briegel1998quantum,kalb2017entanglement}, which are important for quantum networking. 
With a natural abundance of 1.1\%, \nuc nuclei are an easily accessible resource providing a range of possible coupling strengths~\cite{nizovtsev2018non}.
While initialization and coherent control of \nuc spins was already demonstrated for the \siv~\cite{metsch19initialization,nguyen2019quantum,maity2022mechanical,stas2022robust}, this is an ongoing challenge for heavier group-IV defects~\cite{beukers2024control}, that can operate at elevated temperatures without loss of coherence~\cite{debroux2021quantum,rosenthal2023microwave, guo23microwave,wang2024transform}. \\
In this work, we focus on a multi-qubit node comprising a negatively charged germanium-vacancy (GeV) and a strongly coupled \nuc nucleus.
We characterize the hyperfine interaction through optically detected magnetic resonance (ODMR) measurements and a detailed numerical model.
Because of the strong hyperfine interaction, we can initialize the nuclear spin up to 95\% utilizing a power-efficient microwave spin pumping scheme.
Additionally, narrow-band microwave (MW) and radiofrequency (RF) pulses enable conditional flipping of the electron and nuclear spin.
This allows us to efficiently swap the electron and nuclear eigenstates, leading to a projection-SWAP gate, that can be used for initialization and readout.
We demonstrate coherent control of the \nuc spin using direct RF manipulation, achieving nuclear Rabi frequencies up to 11.73\,kHz.
Through application of dynamical decoupling (DD) sequences, we can prolong the nuclear spin's dephasing time of $T_2^* = 2\text{\,ms}$ by more than three orders of magnitude to $T_{2}=2.57\text{\,s}$, marking the longest achieved coherence time of a controllable nuclear memory spin for all group-IV defects~\cite{knaut2024entanglement}. 
We further provide a realistic estimation of the longest achievable memory time of $T^{\text{mem}}_2=18.1\,\text{s}$ considering noise errors modeled by Ornstein-Uhlenbeck (OU) processes~\cite{uhlenbeck1930on,senkalla2024germanium}, the $T_1$ time of the GeV electron spin, and experimental heating limitations.
%

	\textit{The System.--- }
%
\begin{figure}
    \centering
    \includegraphics[width=8.6 cm]{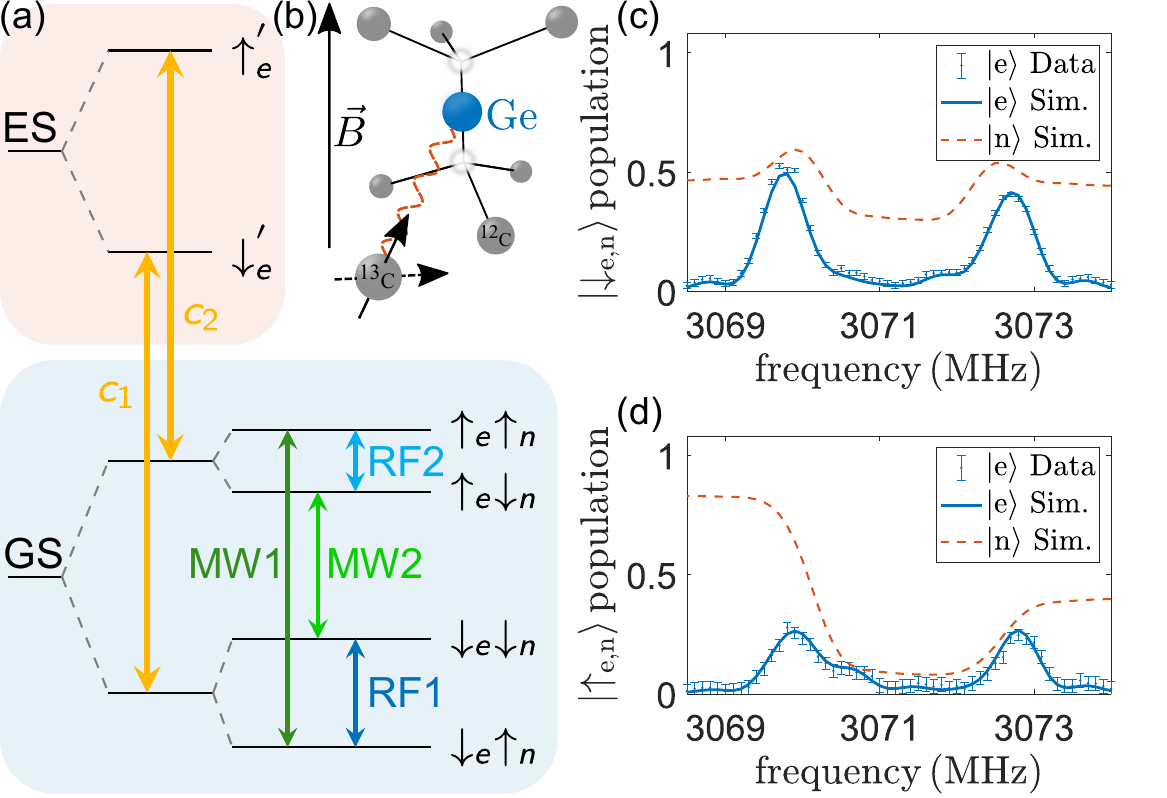} 
    \caption{Description of the system.\quad(a) Reduced energy level scheme of the GeV showing the lower orbital branch in the ground state (GS) and excited state (ES). 
    Strong hyperfine coupling to a \nuc nuclear spin leads to an additional splitting. 
    (b) Schematic of GeV aligned with external magnetic field, showing electron spin-state-dependent nuclear spin quantization axes.
    Pulsed ODMR measurements with descending frequency sweep direction where the electron is reinitialized into $\ket{\uparrow_e}$ via optical transition $c_1$ (c) and into $\ket{\downarrow_e}$ via $c_2$ (d). 
    The red dashed line corresponds to the simulated nuclear spin evolution.}
    \label{fig:fig1}
\end{figure}
We perform the experiments on a High Pressure High Temperature (HPHT) grown diamond with natural abundance of \nucc. 
Our multi-qubit node features a naturally incorporated \gev strongly coupled to a \nuc nuclear spin, operated in a dilution refrigerator 
\cite{supplement,senkalla2024germanium}.
In a magnetic field aligned to the defect's axis,
this forms a four-level system with selective transitions as depicted in Fig. \ref{fig:fig1}\,(a).
The electron (nuclear) spin can be coherently controlled via MW (RF) driving fields, supplied through a 20\,\textmu m thick copper wire spanned over the diamond.
Resonant addressing at the optical transitions $c_{1,2}$ allows for high fidelity initialization and readout of the electron spin.
Further details of the experimental implementation can be found in Ref. \cite{supplement}.\\
%
%
%
The evolution of the system can be characterized in the rotating frame of the \gev center's electron spin by the Hamiltonian 
\begin{equation}
    H = \Delta S_z + \Omega_e S_x - \gamma_n B_z I_z + A_{zx} S_z I_x + A_{zz} S_z I_z,
    \label{eq:H}
\end{equation}
under the secular approximation \cite{maze2012free}, excluding the nuclear RF driving fields for readability.  
The parameters $A_{zx}$ and $A_{zz}$ correspond to the hyperfine coupling parameters, $S_j (I_j)$ with $j\in \{x,y,z\}$ are the spin operators on the GeV electron spin ($^{13}$C nuclear spin), $\gamma_n$ is the gyromagnetic ratio of the nuclear spin, $\Omega_e$ is the Rabi frequency on the electron spin transition and $\Delta$ is the detuning of the MW driving field frequency from the transition frequency of the bare GeV electron spin, i.e., in the absence of $^{13}$C couplings (in angular frequency units)~\cite{supplement}.\\
The strong hyperfine interaction leads to an additional splitting of the electron spin states that can be observed in a pulsed ODMR measurement.
Figure~\ref{fig:fig1}\,(c) shows the spectrum for $\ket{\uparrow_e}$ and Fig.~\ref{fig:fig1}\,(d) for $\ket{\downarrow_e}$, with $\Omega_e=(2\pi)\,349\,\text{kHz}$ for the descending frequency sweep direction.
The data corresponds to the averaged fluorescence signal of $\sim3470$ repetitions where we perform a consecutive Rabi measurement after each repetition for normalization~\cite{supplement}.
We observe a significant difference in shape and amplitude of the spectra, that is heavily influenced by the frequency sweep direction~\cite{supplement}. \\
To identify the underlying dynamics we perform a total of four ODMR measurements, two for each electron spin state, differing by their frequency sweep direction and fit them with a numerical model.
The model describes the full interaction between the electron and nuclear spin, including dephasing and the optical reset of the electron spin. 
Full details can be found in the Supplementaly Material~\cite{supplement}.
The fit result is shown in Fig.~\ref{fig:fig1}\,(c) for $\ket{\uparrow_e}$ and in Fig.~\ref{fig:fig1}\,(d) for $\ket{\downarrow_e}$ by the blue lines, achieving a $R^2$ of 0.9854 and 0.9568, respectively. 
From the fit result we determine the magnetic field to be $B_z=(97.159\pm0.005)\,\text{mT}$ which  slightly deviates from the set 100\,mT due to the diamond placement in the vector magnet.
Using the nuclear transition frequencies $\omega_{\text{RF2}}=(2\pi)\,(493.62\pm0.04)\,\text{kHz}$ and $\omega_{\text{RF1}}=(2\pi)\,(2489.73\pm0.06)\,\text{kHz}$ from a nuclear Ramsey experiment (detailed below) we can estimate the hyperfine coupling parameters according to
\begin{equation}
    A_{zz} = \frac{\omega_{\text{RF1}}^2 - \omega_{\text{RF2}}^2}{2\gamma_nB_z}
\label{eq:A_zz}
\end{equation}
and
\begin{equation}
    A_{zx} = \sqrt{4 \omega_{\text{RF2}}^2 - \left(A_{zz} - 2\gamma_nB_z\right)^2}.
\label{eq:A_zx}
\end{equation}
This leads to $A_{zx}=2\pi\left(602.81\pm0.27\right)\text{\,kHz}$ and $A_{zz}=2\pi\left(2862.34\pm0.14\right)\text{\,kHz}$. \\
Depending on the electron spin state, the strong hyperfine interaction changes the nuclear spin's quantization axis significantly, as illustrated in Fig.~\ref{fig:fig1}\,(b).
Rotation of the electron spin, e.g., in the form of a narrow-band (detuned) $\pi$ pulse, can thus alter the nuclear spin state, which in turn affects the electron spin population transfer.
During a free evolution time, i.e., during the 5.5\,ms long reinitialization of the electron spin, the nuclear spin oscillates around the corresponding quantization axis and dephases. 
Its simulated evolution over the course of the ODMR measurement is shown by the dashed red line, revealing a significant polarization buildup in the $\ket{\downarrow_e}$ manifold [Fig.~\ref{fig:fig1}\,(d)] due to spin pumping. \\
The spin pumping is facilitated by the strong hyperfine interaction that leads to the large splitting $\omega_{\text{RF1}}$ of the nuclear spin states for $\ket{\downarrow_e}$.
This splitting allows the microwave pulse to selectively address one of the hyperfine levels while leaving the other untouched.
The strong $A_{zx}$ coupling then enables the simultaneous flipping of both spins allowing for the otherwise forbidden transitions~\cite{EPR_jeschke}.
As the laser resets the electron spin, the repetitive application of a microwave and laser pulse leads to polarization of the nuclear spin.
At sub-Kelvin temperatures, the nuclear spin's polarization survives the entire sequence and thus the frequency sweep direction alters the observed shape.
For $\ket{\uparrow_e}$ shown in Fig.~\ref{fig:fig1}\,(c), the polarization buildup is significantly less efficient as the small splitting $\omega_{\text{RF2}}$ prevents the selective addressing of the hyperfine levels with the chosen Rabi frequency. \\
We can exploit the spin pumping process to experimentally initialize the nuclear spin for $\ket{\downarrow_e}$ with a state preparation fidelity of 95\,\% by repeatedly applying a resonant MW pulse at $\omega_{\text{MW2}}$ and a subsequent laser pulse 15 times, closely matching our simulation results~\cite{supplement}.
While this method is less time-efficient compared to the projection-SWAP gate shown in the following section, it achieves nuclear spin polarization without the need of resonant RF pulses. Thus, it offers an easily accessible way of initializing the system to its eigenstates while reducing the heat load. 
Additionally, the application of optimal control~\cite{d2021introduction, muller2022one, vetter2024gate} in our numerical simulation suggests, that a fidelity of up to $99$~\% could be achieved, using a single optimized pulse~\cite{supplement}.
\begin{figure}
    \centering
    \includegraphics[width=8.6 cm]{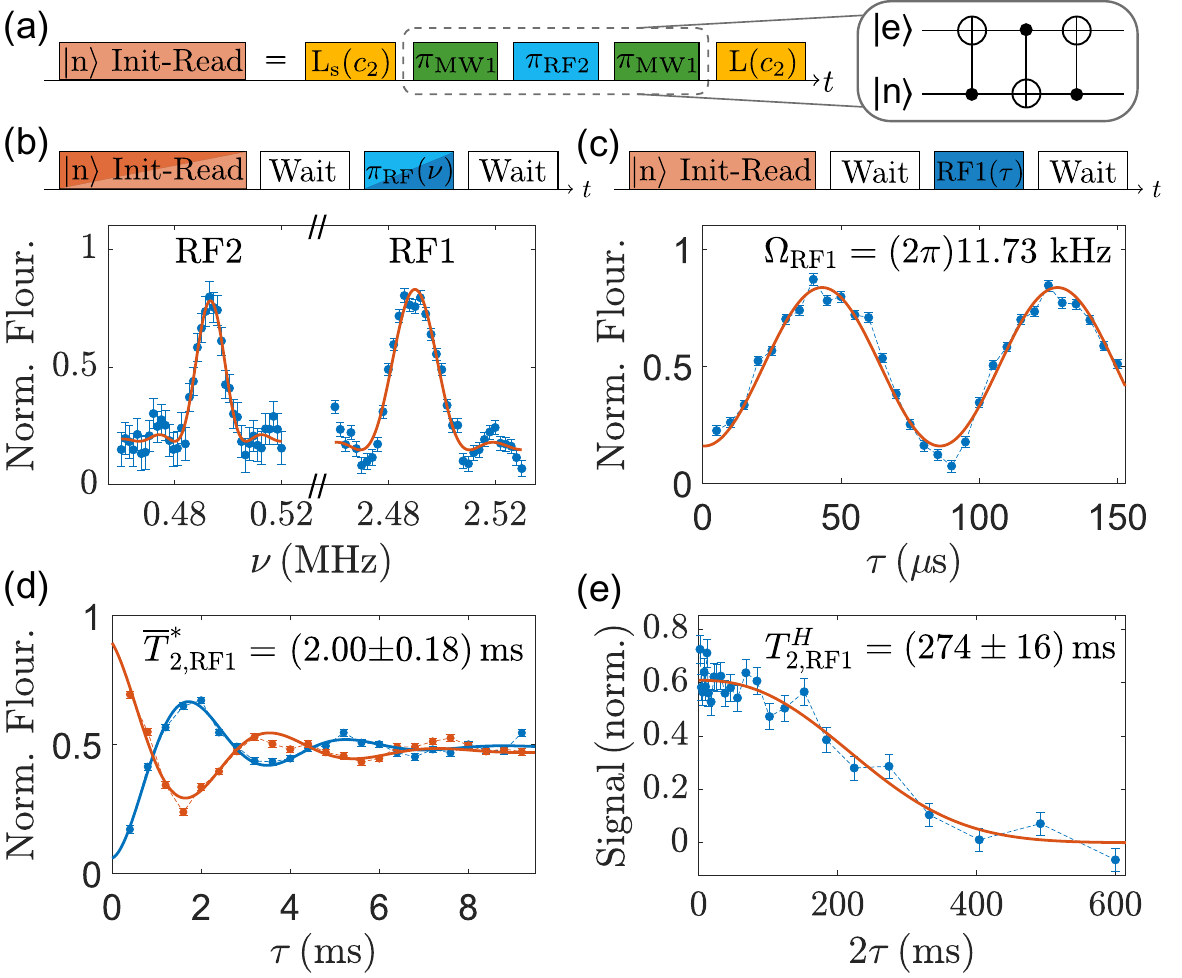}
    \caption{Coherent control of the \nuc nuclear spin.\quad(a) Pulse sequence for nuclear spin initialization and readout in the $\ket{\downarrow_e}$ manifold, where $\mathrm{L_s}$ indicates an initializing laser element without recording the corresponding photons~\cite{supplement}.
    (b) Pulse sequence and measurements for detecting nuclear transition frequencies, with added waiting periods to limit heat load. Nuclear spin initialization and readout in the $\ket{\uparrow_e}$ manifold (dark orange) uses laser frequency $c_1$ and microwave elements MW2 instead.
    The following nuclear spin measurements are conducted in the $\ket{\downarrow_e}$ manifold:
    (c) Demonstration of coherent control by the \nuc Rabi experiment.
    (d) Nuclear Ramsey measurement, with the red line showing a Ramsey measurement using a $180^\circ$ phase-shifted final $\pi/2$ RF pulse.
    (e) Differential signal of a nuclear Hahn echo with corresponding fit (red line).}
    \label{fig:fig2}
\end{figure}

\textit{Coherent control of the nuclear memory spin.--- }
%
To establish coherent control over our nuclear spin, we construct a projection-SWAP gate based on three CNOT gates, as sketched in Fig.~\ref{fig:fig2}\,(a).
The large $A_{zz}$ hyperfine coupling allows us to implement these CNOT gates up to a phase through narrow-band $\pi$-pulse gates that flip the electron spin conditional on the nuclear spin and vice versa. 
The limited electron dephasing time of $T_{2,e}^*=1.43\text{\,\textmu s}$ \cite{senkalla2024germanium} poses challenges for a direct transfer of coherent states, given the dozens of microseconds long $\text{C}_e\text{NOT}_n$ gate.
Nonetheless, we can use the projection-SWAP gate to efficiently initialize the nuclear spin to its eigenstates and read out the projection of any arbitrary nuclear spin state onto $\ket{\uparrow_n}$ or $\ket{\downarrow_n}$. \\
The projection-SWAP gate requires knowledge of the nuclear transition frequency and $\pi$ RF pulse length of the $\text{C}_e\text{NOT}_n$ gate. 
To determine these parameters the RF pulse can be sandwiched in between the previously introduced nuclear spin pumping scheme for initialization and a $\text{C}_n\text{NOT}_e$ combined with a subsequent laser pulse for readout.
Alternatively, the required parameters can be obtained by iteratively maximizing the contrast of the RF-frequency ($\nu$) sweep experiment in Fig.~\ref{fig:fig2}\,(b) and the RF-Rabi experiment in Fig.~\ref{fig:fig2}\,(c).
We obtain a nuclear Rabi frequency of $(11.73\pm0.14)\text{\,kHz}$ for $\omega_{\text{RF1}}$, which we can use to construct any arbitrary single qubit nuclear spin gate, establishing full coherent control over the nuclear subspace. \\
To characterize the quality of our nuclear memory storage, we start by measuring the nuclear dephasing time in a Ramsey experiment, consisting of two $\pi/2$ RF pulses with a variable interpulse duration.
Figure~\ref{fig:fig2}\,(d) illustrates that measurements are performed in an alternating manner, using a final $\pi/2$ RF pulse with $0^\circ$ ($x$) or $180^\circ$ ($-x$) phase to project onto the different eigenstates.
As this helps to minimize effects like laser fluctuations, all of the following measurements are performed in this fashion, while only the normalized differential signal is shown.
The oscillations of the signal shown in Fig.~\ref{fig:fig2}\,(d) arise from controlled detuning of the applied field from the nuclear transitions, which allows us to infer their frequencies with high accuracy to $\omega_{\text{RF1}}=(2\pi)(2489.73\pm0.06)\,\text{kHz}$ and $\omega_{\text{RF2}}=(2\pi)(493.62\pm0.04)\,\text{kHz}$ \cite{supplement}. \\
Through application of a Hahn echo sequence~\cite{hahn1950spin} we can decouple the nuclear spin from slowly varying magnetic fields, resulting in a coherence time of $T_2^H = (274\pm 16)\,\text{ms}$, as displayed in Fig. \ref{fig:fig2}(e) -- an improvement of more than two orders of magnitude compared to $T_{2,\text{RF1}}^*=(2.00\pm 0.18)\,\text{ms}$ obtained by the Ramsey measurement.

\textit{Analysis of the nuclear spin quantum memory.--- }
\begin{figure}
    \centering
    \includegraphics[width=8.6 cm]{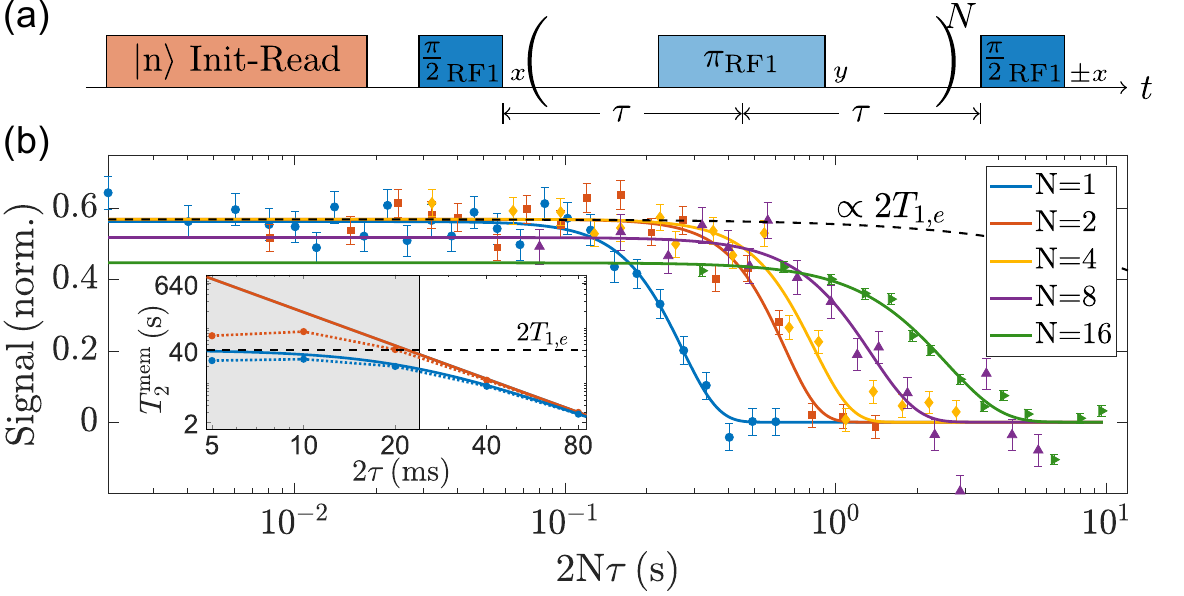}
    \caption{Coherence time measurements on \nucc.\quad(a) Pulse sequence of the nuclear CPMG sequence with the number of refocusing pulses $N$ and interpulse spacing $\tau$ in alternating manner (indicated by $\pm x$ on final $\pi/2$ RF pulse). 
    (b)	Corresponding coherence time measurements for fixed $N$ with their respective fit. 
    The black dashed line represents the theoretical limit of the electron relaxation time $T_{1,e}$. 
    Inset: Simulation of the limit of the expected memory time when using different pulse separation times $2\tau$, assuming an OU noise process. 
    The blue (red) line shows the expected memory time with (without) taking into account $T_{1,e}$ relaxation.
    The white area emphasizes feasible pulse spacings that avoid excessive heating in our current setup. 
    The dotted curves show numerical simulations of the memory time with the XY8 sequence that take pulse errors into account~\cite{supplement}.}
    \label{fig:fig3}
\end{figure}
The $^{13}$C nuclear spin coherence time can be further extended by dynamical decoupling protocols \cite{viola1998dynamical, meiboom1958modified, gullion1990new, uhrig2007keeping, ryan2010robust, souza2011robust}.
We apply the CPMG sequence~\cite{meiboom1958modified} for our experiments, similar to previous work \cite{senkalla2024germanium}, as it can also be used to characterize the environmental noise.
For this purpose, we start by measuring the coherence time for different orders $N$ (equal to the number of refocusing pulses), while sweeping the pulse spacing $\tau$.
The corresponding sequence is shown in Fig.~\ref{fig:fig3}\,(a).
We achieve a coherence time up to $T_2 =(2.57\pm0.19)\,\text{s}$ for $N=16$, shown in green in Fig. \ref{fig:fig3}(b), which marks the so far longest measured nuclear spin coherence time for any group-IV defect.\\
The overall contrast decays with increasing $N$, which can be explained by the insufficient long-term stability of the laser hardware over the extended measurement period ($>$1\,d for $N=16$) or by minor pulse imperfections.
To investigate the impact of pulse errors, we perform an additional measurement using the XY8 sequence, known for its robustness \cite{gullion1990new}. 
The resulting coherence time of $T_2^{\text{XY8}}=(1.49\pm0.10)\,\text{s}$ \cite{supplement} is consistent with the CPMG-8 sequence, $T_2^{\text{CPMG8}}=(1.36\pm0.14)\,\text{s}$, suggesting that pulse errors are not yet limiting our coherence time. \\
Building on these results, we assess the memory capabilities of our system.
We model the effects of environmental and amplitude noise using an Ornstein-Uhlenbeck process, as outlined in previous work \cite{senkalla2024germanium} and further elaborated in Ref. ~\cite{supplement}.
The inset of Fig.~\ref{fig:fig3}(b) shows the possible \nuc memory time, using a conservative estimate of the noise correlation time \cite{supplement}, and accounting for pulse errors and heating limitations in our setup.
The main limiting factors for the memory time are heating and $T_{1,e}$ relaxation.  
We estimate that a pulse spacing of $2\tau=24$\,ms and higher causes negligible heating and is feasible for dynamical decoupling. This allows for a memory time of $T^{\text{mem}}_2= 18.1\,$s \cite{supplement}. 
Reducing pulse separation to around 10\,ms by improving the effective cooling of the sample would boost the memory time to $T^{\text{mem}}_2\approx 28\,$s where pulse errors with the XY8 sequence start to play a role due to the large number of pulses.
Application of higher order sequences like KDD or the UR family~\cite{Souza2012,Genov2017PRL,Ezzell23} can in principle boost the memory time limit even closer to the $T_{1,e}$ limit of $41.4\,$s. 

\textit{Discussion.--- }
%
The proximity of a \nuc nuclear spin to the \gev center within our multi-qubit node results in a strong $A_{zx}$ and $A_{zz}$ hyperfine coupling between the nuclear and electron spin. 
While this influences fundamental pulsed experiments, we show how to leverage it to determine the system's Hamiltonian with its hyperfine parameters by providing an elaborate numerical model which can be fitted to a pulsed ODMR measurement.
\\
We use the insights of the model to implement a cooling power-efficient spin pumping scheme to initialize the \nuc nuclear spin with the so far highest fidelity of 95\% utilizing a group-IV defect.
This approach is particularly advantageous as it does not require precise prior knowledge of the coupling parameters and addresses the constraints of cooling power in dilution refrigerators.
Although this protocol takes longer compared to the demonstrated projection-SWAP gate, quantum optimal control~\cite{rossignolo_quocs_2023, d2021introduction, muller2022one, vetter2024gate} offers a potential path to even shorter gate times and higher fidelities~\cite{supplement}. 
Our spin pumping scheme and numerical model can be seamlessly applied to any group-IV color center with $S=1/2$, provided it strongly interacts with a nonaxial nuclear $I=1/2$ spin.\\
Direct driving of the \nuc nuclear spin with resonant RF pulses allows precise one-qubit operations, bypassing the limitations of using the electron spin as an intermediary \cite{nguyen2019quantum,maity2022mechanical}, and facilitates straightforward implementation of a $\text{C}_e\text{NOT}_n$.
The strong $A_{zz}$ coupling enables the implementation of a $\text{C}_n\text{NOT}_e$ in a similar fashion through a narrow-band microwave pulse. 
Combining both of these gates allows to easily implement a projection-SWAP that can be used to initialize the nuclear spin and readout the projection of its state along its quantization axis.
In its current form, our projection-SWAP gate cannot swap coherent states as the $\text{C}_e\text{NOT}_n$ length significantly exceeds the electron's dephasing time.
While protocols that account for the low dephasing time by interleaving the gate with refocusing pulses~\cite{zhang2015experimental, stas2022robust} are hindered by our strong $A_{zx}$ interaction, advanced protocols~\cite{vallabhapurapu2022indirect, zhang2019improved, hegde2020efficient} or quantum optimal control might also provide a solution to overcome this problem~\cite{zhang2023coupling, dolde2014high}.

\textit{Conclusion and outlook.--- }
We demonstrate coherent control over the nuclear spin of $^{13}$C, strongly coupled to a GeV center electron spin. 
We perform dynamical decoupling to protect the nuclear spin from environmental noise for more than 2.5\,\text{s}, marking the so far longest achieved nuclear memory time for all group-IV defect systems. 
This is achieved with a 60 times lower number of refocusing pulses than previously required for coherence times beyond 2\,s for group-IV defects~\cite{stas2022robust}.
The low duty cycle results in overall reduced heat load, which is critical for operation in dilution refrigerators.
To estimate the memory performance, we include these heating effects, electron spin $T_{1,e}$ relaxation, pulse errors and environmental noise simulated by Ornstein-Uhlenbeck processes in a numerical model.
We can thus provide a realistic upper limit of $18.1\,\text{s}$ as an achievable memory time with our current setup, which can realistically be extended to several tens of seconds through modifications, such as optimizing thermal grounding, improving the MW and RF delivery or using superconducting structures~\cite{karapatzakis2024microwavecontroltinvacancyspin}.\\
Such a long living quantum memory facilitates the generation of multidimensional cluster states \cite{Michaels2021multidimensional}, which are important in the realm of measurement based quantum computing or long distance quantum communication.
Furthermore, integrating our multi-qubit node into a nanophotonic cavity could facilitate the implementation of recently demonstrated direct photon-nuclear entanglement gates~\cite{stas2022robust}, which have achieved successful entanglement generation over distances of 35\,km~\cite{knaut2024entanglement}.
Finally, leveraging weakly coupled \nuc spins, which can be actively decoupled from the electron spin, would allow for scaling up the register \cite{taminiau2014universal,bradley2019ten,van2024mapping,zahedian24blueprint,joas2024high, beukers2024control} and pave the way for performing QEC~\cite{taminiau2014universal, unden2016quantum} and entanglement purification protocols~\cite{briegel1998quantum,kalb2017entanglement}.
\\
\begin{acknowledgments}
\textit{Acknowledgment.--- }
We thank Frank Wilhelm-Mauch, R\'emi Blinder, Lev Kazak, and Petr Siyushev for helpful discussions and technical support. Further, we thank Yuri N. Palyanov, Igor N. Kupriyanov and Yuri M. Borzdov for providing the sample used in this work. The authors acknowledge support by the state of Baden-Württemberg through bwHPC.\\
This work was funded by the German Federal Ministry of Research (BMBF) by future cluster QSENS (No. 03ZK110AB) and projects DE-Brill (No. 13N16207), SPINNING (No. 13N16210 and No. 13N16215), DIAQNOS (No. 13N16463), Quamapolis (No. 13N15375), and EXTRASENS (No. 731473 and 101017733) DLR via project QUASIMODO (No. 50WM2170), Deutsche Forschungsgemeinschaft (DFG) via Projects No. 386028944, No. 445243414, No. 491245864, and No. 387073854.
This project has also received funding from the European Union’s HORIZON Europe program via projects QuMicro (No. 101046911), SPINUS (No. 101135699), C-QuENS (No. 101135359), QCIRCLE (No. 101059999) and FLORIN (No. 101086142), European Research Council (ERC) via Synergy grant HyperQ (No. 856432), Competence Center Quantum Computing Baden-Württemberg, and Carl-Zeiss-Stiftung via the Center of Integrated Quantum Science and technology (IQST) and project Utrasens-Vir (No. P2022-06-007) and QPhoton, as well as the Helmholtz Validation Fund project “Qruise” (HVF-00096).

\end{acknowledgments}
\newpage
\bibliographystyle{ieeetr}
\bibliography{references}

\begin{thebibliography}{10}

\bibitem{nickerson2014freely}
N.~H. Nickerson, J.~F. Fitzsimons, and S.~C. Benjamin, ``Freely scalable
  quantum technologies using cells of 5-to-50 qubits with very lossy and noisy
  photonic links,'' {\em Phys. Rev. X}, vol.~4, p.~041041, Dec 2014.

\bibitem{choi2019percolation}
H.~Choi, M.~Pant, S.~Guha, and D.~Englund, ``Percolation-based architecture for
  cluster state creation using photon-mediated entanglement between atomic
  memories,'' {\em npj Quantum Information}, vol.~5, no.~1, p.~104, 2019.

\bibitem{cuomo2020towards}
D.~Cuomo, M.~Caleffi, and A.~S. Cacciapuoti, ``Towards a distributed quantum
  computing ecosystem,'' {\em IET Quantum Communication}, vol.~1, no.~1,
  pp.~3--8, 2020.

\bibitem{drmota2024verifiable}
P.~Drmota, D.~P. Nadlinger, D.~Main, B.~C. Nichol, E.~M. Ainley, D.~Leichtle,
  A.~Mantri, E.~Kashefi, R.~Srinivas, G.~Araneda, C.~J. Ballance, and D.~M.
  Lucas, ``Verifiable blind quantum computing with trapped ions and single
  photons,'' {\em Phys. Rev. Lett.}, vol.~132, p.~150604, Apr 2024.

\bibitem{ekert1991quantum}
A.~K. Ekert, ``Quantum cryptography based on bell's theorem,'' {\em Phys. Rev.
  Lett.}, vol.~67, pp.~661--663, Aug 1991.

\bibitem{elliott2002building}
C.~Elliott, ``Building the quantum network,'' {\em New Journal of Physics},
  vol.~4, no.~1, p.~46, 2002.

\bibitem{lo2014secure}
H.-K. Lo, M.~Curty, and K.~Tamaki, ``Secure quantum key distribution,'' {\em
  Nature Photonics}, vol.~8, no.~8, pp.~595--604, 2014.

\bibitem{xu2020secure}
F.~Xu, X.~Ma, Q.~Zhang, H.-K. Lo, and J.-W. Pan, ``Secure quantum key
  distribution with realistic devices,'' {\em Reviews of modern physics},
  vol.~92, no.~2, p.~025002, 2020.

\bibitem{gottesman2012longer}
D.~Gottesman, T.~Jennewein, and S.~Croke, ``Longer-baseline telescopes using
  quantum repeaters,'' {\em Phys. Rev. Lett.}, vol.~109, p.~070503, Aug 2012.

\bibitem{komar2014quantum}
P.~Komar, E.~M. Kessler, M.~Bishof, L.~Jiang, A.~S. S{\o}rensen, J.~Ye, and
  M.~D. Lukin, ``A quantum network of clocks,'' {\em Nature Physics}, vol.~10,
  no.~8, pp.~582--587, 2014.

\bibitem{khabiboulline2019optical}
E.~T. Khabiboulline, J.~Borregaard, K.~De~Greve, and M.~D. Lukin, ``Optical
  interferometry with quantum networks,'' {\em Phys. Rev. Lett.}, vol.~123,
  p.~070504, Aug 2019.

\bibitem{yang2024quantum}
Y.~Yang, B.~Yadin, and Z.-P. Xu, ``Quantum-enhanced metrology with network
  states,'' {\em Phys. Rev. Lett.}, vol.~132, p.~210801, May 2024.

\bibitem{briegel1998quantum}
H.-J. Briegel, W.~D\"ur, J.~I. Cirac, and P.~Zoller, ``Quantum repeaters: The
  role of imperfect local operations in quantum communication,'' {\em Phys.
  Rev. Lett.}, vol.~81, pp.~5932--5935, Dec 1998.

\bibitem{ruf2021quantum}
M.~Ruf, N.~H. Wan, H.~Choi, D.~Englund, and R.~Hanson, ``Quantum networks based
  on color centers in diamond,'' {\em Journal of Applied Physics}, vol.~130,
  no.~7, 2021.

\bibitem{rogers2014alloptical}
L.~J. Rogers, K.~D. Jahnke, M.~H. Metsch, A.~Sipahigil, J.~M. Binder,
  T.~Teraji, H.~Sumiya, J.~Isoya, M.~D. Lukin, P.~Hemmer, and F.~Jelezko,
  ``All-optical initialization, readout, and coherent preparation of single
  silicon-vacancy spins in diamond,'' {\em Phys. Rev. Lett.}, vol.~113,
  p.~263602, Dec 2014.

\bibitem{siyushev2017optical}
P.~Siyushev, M.~H. Metsch, A.~Ijaz, J.~M. Binder, M.~K. Bhaskar, D.~D.
  Sukachev, A.~Sipahigil, R.~E. Evans, C.~T. Nguyen, M.~D. Lukin, P.~R. Hemmer,
  Y.~N. Palyanov, I.~N. Kupriyanov, Y.~M. Borzdov, L.~J. Rogers, and
  F.~Jelezko, ``Optical and microwave control of germanium-vacancy center spins
  in diamond,'' {\em Phys. Rev. B}, vol.~96, p.~081201(R), Aug 2017.

\bibitem{gorlitz2020spectroscopic}
J.~G{\"o}rlitz, D.~Herrmann, G.~Thiering, P.~Fuchs, M.~Gandil, T.~Iwasaki,
  T.~Taniguchi, M.~Kieschnick, J.~Meijer, M.~Hatano, {\em et~al.},
  ``Spectroscopic investigations of negatively charged tin-vacancy centres in
  diamond,'' {\em New Journal of Physics}, vol.~22, no.~1, p.~013048, 2020.

\bibitem{wang2024transform}
P.~Wang, L.~Kazak, K.~Senkalla, P.~Siyushev, R.~Abe, T.~Taniguchi, S.~Onoda,
  H.~Kato, T.~Makino, M.~Hatano, F.~Jelezko, and T.~Iwasaki,
  ``Transform-limited photon emission from a lead-vacancy center in diamond
  above 10 k,'' {\em Phys. Rev. Lett.}, vol.~132, p.~073601, Feb 2024.

\bibitem{bradac2019quantum}
C.~Bradac, W.~Gao, J.~Forneris, M.~E. Trusheim, and I.~Aharonovich, ``Quantum
  nanophotonics with group {IV} defects in diamond,'' {\em Nature
  communications}, vol.~10, no.~1, p.~5625, 2019.

\bibitem{sipahigil2016integrated}
A.~Sipahigil, R.~E. Evans, D.~D. Sukachev, M.~J. Burek, J.~Borregaard, M.~K.
  Bhaskar, C.~T. Nguyen, J.~L. Pacheco, H.~A. Atikian, C.~Meuwly, {\em et~al.},
  ``An integrated diamond nanophotonics platform for quantum-optical
  networks,'' {\em Science}, vol.~354, no.~6314, pp.~847--850, 2016.

\bibitem{bhaskar2017quantum}
M.~K. Bhaskar, D.~D. Sukachev, A.~Sipahigil, R.~E. Evans, M.~J. Burek, C.~T.
  Nguyen, L.~J. Rogers, P.~Siyushev, M.~H. Metsch, H.~Park, F.~Jelezko,
  M.~Lon\ifmmode~\check{c}\else \v{c}\fi{}ar, and M.~D. Lukin, ``Quantum
  nonlinear optics with a germanium-vacancy color center in a nanoscale diamond
  waveguide,'' {\em Phys. Rev. Lett.}, vol.~118, p.~223603, May 2017.

\bibitem{rugar2021quantum}
A.~E. Rugar, S.~Aghaeimeibodi, D.~Riedel, C.~Dory, H.~Lu, P.~J. McQuade, Z.-X.
  Shen, N.~A. Melosh, and J.~Vu\ifmmode \check{c}\else
  \v{c}\fi{}kovi\ifmmode~\acute{c}\else \'{c}\fi{}, ``Quantum photonic
  interface for tin-vacancy centers in diamond,'' {\em Phys. Rev. X}, vol.~11,
  p.~031021, Jul 2021.

\bibitem{antoniuk2024all}
L.~Antoniuk, N.~Lettner, A.~P. Ovvyan, S.~Haugg, M.~Klotz, H.~Gehring,
  D.~Wendland, V.~N. Agafonov, W.~H. Pernice, and A.~Kubanek, ``All-optical
  spin access via a cavity-broadened optical transition in on-chip hybrid
  quantum photonics,'' {\em Phys. Rev. Appl.}, vol.~21, p.~054032, May 2024.

\bibitem{parker2024diamond}
R.~A. Parker, J.~Arjona~Mart{\'\i}nez, K.~C. Chen, A.~M. Stramma, I.~B. Harris,
  C.~P. Michaels, M.~E. Trusheim, M.~Hayhurst~Appel, C.~M. Purser, W.~G. Roth,
  {\em et~al.}, ``A diamond nanophotonic interface with an optically accessible
  deterministic electronuclear spin register,'' {\em Nature Photonics},
  vol.~18, no.~2, pp.~156--161, 2024.

\bibitem{riegel24lifetimereduction}
R.~Zifkin, C.~D. Rodr\'{\i}guez~Rosenblueth, E.~Janitz, Y.~Fontana, and
  L.~Childress, ``Lifetime reduction of single germanium-vacancy centers in
  diamond via a tunable open microcavity,'' {\em PRX Quantum}, vol.~5,
  p.~030308, Jul 2024.

\bibitem{pingault2017coherent}
B.~Pingault, D.-D. Jarausch, C.~Hepp, L.~Klintberg, J.~N. Becker, M.~Markham,
  C.~Becher, and M.~Atat{\"u}re, ``Coherent control of the silicon-vacancy spin
  in diamond,'' {\em Nature communications}, vol.~8, no.~1, p.~15579, 2017.

\bibitem{harris2024coherence}
I.~B.~W. Harris and D.~Englund, ``Coherence of group-{IV} color centers,'' {\em
  Phys. Rev. B}, vol.~109, p.~085414, Feb 2024.

\bibitem{meesala18strainengineering}
S.~Meesala, Y.-I. Sohn, B.~Pingault, L.~Shao, H.~A. Atikian, J.~Holzgrafe,
  M.~G\"undo\ifmmode~\breve{g}\else \u{g}\fi{}an, C.~Stavrakas, A.~Sipahigil,
  C.~Chia, R.~Evans, M.~J. Burek, M.~Zhang, L.~Wu, J.~L. Pacheco, J.~Abraham,
  E.~Bielejec, M.~D. Lukin, M.~Atat\"ure, and M.~Lon\ifmmode~\check{c}\else
  \v{c}\fi{}ar, ``Strain engineering of the silicon-vacancy center in
  diamond,'' {\em Phys. Rev. B}, vol.~97, p.~205444, May 2018.

\bibitem{sohn2018controlling}
Y.-I. Sohn, S.~Meesala, B.~Pingault, H.~A. Atikian, J.~Holzgrafe,
  M.~G{\"u}ndo{\u{g}}an, C.~Stavrakas, M.~J. Stanley, A.~Sipahigil, J.~Choi,
  {\em et~al.}, ``Controlling the coherence of a diamond spin qubit through its
  strain environment,'' {\em Nature communications}, vol.~9, no.~1, p.~2012,
  2018.

\bibitem{stas2022robust}
P.-J. Stas, Y.~Q. Huan, B.~Machielse, E.~N. Knall, A.~Suleymanzade,
  B.~Pingault, M.~Sutula, S.~W. Ding, C.~M. Knaut, D.~R. Assumpcao, {\em
  et~al.}, ``Robust multi-qubit quantum network node with integrated error
  detection,'' {\em Science}, vol.~378, no.~6619, pp.~557--560, 2022.

\bibitem{zu2022development}
H.~Zu, W.~Dai, and A.~De~Waele, ``Development of dilution refrigerators—a
  review,'' {\em Cryogenics}, vol.~121, p.~103390, 2022.

\bibitem{sukachev2017silicon}
D.~D. Sukachev, A.~Sipahigil, C.~T. Nguyen, M.~K. Bhaskar, R.~E. Evans,
  F.~Jelezko, and M.~D. Lukin, ``Silicon-vacancy spin qubit in diamond: a
  quantum memory exceeding 10 ms with single-shot state readout,'' {\em
  Physical review letters}, vol.~119, no.~22, p.~223602, 2017.

\bibitem{senkalla2024germanium}
K.~Senkalla, G.~Genov, M.~H. Metsch, P.~Siyushev, and F.~Jelezko, ``Germanium
  vacancy in diamond quantum memory exceeding 20 ms,'' {\em Phys. Rev. Lett.},
  vol.~132, p.~026901, Jan 2024.

\bibitem{karapatzakis2024microwavecontroltinvacancyspin}
I.~Karapatzakis, J.~Resch, M.~Schrodin, P.~Fuchs, M.~Kieschnick, J.~Heupel,
  L.~Kussi, C.~S\"urgers, C.~Popov, J.~Meijer, C.~Becher, W.~Wernsdorfer, and
  D.~Hunger, ``Microwave control of the tin-vacancy spin qubit in diamond with
  a superconducting waveguide,'' {\em Phys. Rev. X}, vol.~14, p.~031036, Aug
  2024.

\bibitem{knaut2024entanglement}
C.~Knaut, A.~Suleymanzade, Y.-C. Wei, D.~Assumpcao, P.-J. Stas, Y.~Huan,
  B.~Machielse, E.~Knall, M.~Sutula, G.~Baranes, {\em et~al.}, ``Entanglement
  of nanophotonic quantum memory nodes in a telecom network,'' {\em Nature},
  vol.~629, no.~8012, pp.~573--578, 2024.

\bibitem{taminiau2014universal}
T.~H. Taminiau, J.~Cramer, T.~van~der Sar, V.~V. Dobrovitski, and R.~Hanson,
  ``Universal control and error correction in multi-qubit spin registers in
  diamond,'' {\em Nature nanotechnology}, vol.~9, no.~3, pp.~171--176, 2014.

\bibitem{unden2016quantum}
T.~Unden, P.~Balasubramanian, D.~Louzon, Y.~Vinkler, M.~B. Plenio, M.~Markham,
  D.~Twitchen, A.~Stacey, I.~Lovchinsky, A.~O. Sushkov, M.~D. Lukin,
  A.~Retzker, B.~Naydenov, L.~P. McGuinness, and F.~Jelezko, ``Quantum
  metrology enhanced by repetitive quantum error correction,'' {\em Phys. Rev.
  Lett.}, vol.~116, p.~230502, Jun 2016.

\bibitem{kalb2017entanglement}
N.~Kalb, A.~A. Reiserer, P.~C. Humphreys, J.~J. Bakermans, S.~J. Kamerling,
  N.~H. Nickerson, S.~C. Benjamin, D.~J. Twitchen, M.~Markham, and R.~Hanson,
  ``Entanglement distillation between solid-state quantum network nodes,'' {\em
  Science}, vol.~356, no.~6341, pp.~928--932, 2017.

\bibitem{nizovtsev2018non}
A.~P. Nizovtsev, S.~Y. Kilin, A.~L. Pushkarchuk, V.~A. Pushkarchuk, S.~A.
  Kuten, O.~A. Zhikol, S.~Schmitt, T.~Unden, and F.~Jelezko, ``Non-flipping
  $^{13}${C} spins near an {NV} center in diamond: hyperfine and spatial
  characteristics by density functional theory simulation of the {C510 [NV]}
  {H252} cluster,'' {\em New Journal of Physics}, vol.~20, no.~2, p.~023022,
  2018.

\bibitem{metsch19initialization}
M.~H. Metsch, K.~Senkalla, B.~Tratzmiller, J.~Scheuer, M.~Kern, J.~Achard,
  A.~Tallaire, M.~B. Plenio, P.~Siyushev, and F.~Jelezko, ``Initialization and
  readout of nuclear spins via a negatively charged silicon-vacancy center in
  diamond,'' {\em Phys. Rev. Lett.}, vol.~122, p.~190503, May 2019.

\bibitem{nguyen2019quantum}
C.~T. Nguyen, D.~D. Sukachev, M.~K. Bhaskar, B.~Machielse, D.~S. Levonian,
  E.~N. Knall, P.~Stroganov, R.~Riedinger, H.~Park,
  M.~Lon\ifmmode~\check{c}\else \v{c}\fi{}ar, and M.~D. Lukin, ``Quantum
  network nodes based on diamond qubits with an efficient nanophotonic
  interface,'' {\em Phys. Rev. Lett.}, vol.~123, p.~183602, Oct 2019.

\bibitem{maity2022mechanical}
S.~Maity, B.~Pingault, G.~Joe, M.~Chalupnik, D.~Assump{\c{c}}{\~a}o,
  E.~Cornell, L.~Shao, and M.~Lon\ifmmode~\check{c}\else \v{c}\fi{}ar,
  ``Mechanical control of a single nuclear spin,'' {\em Phys. Rev. X}, vol.~12,
  p.~011056, Mar 2022.

\bibitem{beukers2024control}
H.~K. Beukers, C.~Waas, M.~Pasini, H.~B. van Ommen, N.~Codreanu, J.~M.
  Brevoord, T.~Turan, M.~Iuliano, Z.~Ademi, T.~H. Taminiau, {\em et~al.},
  ``Control of solid-state nuclear spin qubits using an electron spin-1/2,''
  {\em arXiv preprint arXiv:2409.08977}, 2024.

\bibitem{debroux2021quantum}
R.~Debroux, C.~P. Michaels, C.~M. Purser, N.~Wan, M.~E. Trusheim,
  J.~Arjona~Mart\'{\i}nez, R.~A. Parker, A.~M. Stramma, K.~C. Chen,
  L.~de~Santis, E.~M. Alexeev, A.~C. Ferrari, D.~Englund, D.~A. Gangloff, and
  M.~Atat\"ure, ``Quantum control of the tin-vacancy spin qubit in diamond,''
  {\em Phys. Rev. X}, vol.~11, p.~041041, Nov 2021.

\bibitem{rosenthal2023microwave}
E.~I. Rosenthal, C.~P. Anderson, H.~C. Kleidermacher, A.~J. Stein, H.~Lee,
  J.~Grzesik, G.~Scuri, A.~E. Rugar, D.~Riedel, S.~Aghaeimeibodi, G.~H. Ahn,
  K.~Van~Gasse, and J.~Vu\ifmmode \check{c}\else
  \v{c}\fi{}kovi\ifmmode~\acute{c}\else \'{c}\fi{}, ``Microwave spin control of
  a tin-vacancy qubit in diamond,'' {\em Phys. Rev. X}, vol.~13, p.~031022, Aug
  2023.

\bibitem{guo23microwave}
X.~Guo, A.~M. Stramma, Z.~Li, W.~G. Roth, B.~Huang, Y.~Jin, R.~A. Parker,
  J.~Arjona~Mart\'{\i}nez, N.~Shofer, C.~P. Michaels, C.~P. Purser, M.~H.
  Appel, E.~M. Alexeev, T.~Liu, A.~C. Ferrari, D.~D. Awschalom, N.~Delegan,
  B.~Pingault, G.~Galli, F.~J. Heremans, M.~Atat\"ure, and A.~A. High,
  ``Microwave-based quantum control and coherence protection of tin-vacancy
  spin qubits in a strain-tuned diamond-membrane heterostructure,'' {\em Phys.
  Rev. X}, vol.~13, p.~041037, Nov 2023.

\bibitem{uhlenbeck1930on}
G.~E. Uhlenbeck and L.~S. Ornstein, ``On the theory of the {Brownian} motion,''
  {\em Phys. Rev.}, vol.~36, pp.~823--841, Sep 1930.

\bibitem{supplement}
See Supplemental Material for details on the
  sample and setup, for the numerical simulation for hyperfine parameter
  extraction, nuclear spin pumping, optimal control simulations, further
  details on nuclear spin coherence experiments, noise calibration and
  numerical simulations on nuclear spin memory time with references
  \cite{palyanov2015germanium, nielsen2010quantum, thiering2018ab,
  storn1997differential, EPR_jeschke, Barry2020, rach2015, UhlenbeckRMP1945,
  Gillespie1996AJP, Pascual-WinterPRB2012, Bar-GillNatComm2013, Genov2019MDD,
  Genov2020PRR, Aharon2016NJP,barry2023sensitiveacdcmagnetometry} included
  therein.

\bibitem{palyanov2015germanium}
Y.~N. Palyanov, I.~N. Kupriyanov, Y.~M. Borzdov, and N.~V. Surovtsev,
  ``Germanium: a new catalyst for diamond synthesis and a new optically active
  impurity in diamond,'' {\em Scientific reports}, vol.~5, no.~1, p.~14789,
  2015.

\bibitem{nielsen2010quantum}
M.~A. Nielsen and I.~L. Chuang, {\em Quantum computation and quantum
  information}.
\newblock Cambridge university press, 2010.

\bibitem{thiering2018ab}
G.~Thiering and A.~Gali, ``\emph{Ab Initio} magneto-optical spectrum of
  group-{IV} vacancy color centers in diamond,'' {\em Phys. Rev. X}, vol.~8,
  p.~021063, Jun 2018.

\bibitem{storn1997differential}
R.~Storn and K.~Price, ``Differential evolution--a simple and efficient
  heuristic for global optimization over continuous spaces,'' {\em Journal of
  global optimization}, vol.~11, pp.~341--359, 1997.

\bibitem{EPR_jeschke}
A.~Schweiger and G.~Jeschke, {\em Principles of pulse electron paramagnetic
  resonance}.
\newblock Oxford University Press, 2001.

\bibitem{Barry2020}
J.~F. Barry, J.~M. Schloss, E.~Bauch, M.~J. Turner, C.~A. Hart, L.~M. Pham, and
  R.~L. Walsworth, ``Sensitivity optimization for {NV}-diamond magnetometry,''
  {\em Rev. Mod. Phys.}, vol.~92, p.~015004, Mar 2020.

\bibitem{rach2015}
N.~Rach, M.~M. M\"uller, T.~Calarco, and S.~Montangero, ``Dressing the
  chopped-random-basis optimization: A bandwidth-limited access to the
  trap-free landscape,'' {\em Phys. Rev. A}, vol.~92, p.~062343, Dec 2015.

\bibitem{UhlenbeckRMP1945}
M.~C. Wang and G.~E. Uhlenbeck, ``On the theory of the {Brownian} motion
  {II},'' {\em Rev. Mod. Phys.}, vol.~17, pp.~323--342, Apr 1945.

\bibitem{Gillespie1996AJP}
D.~T. Gillespie, ``The mathematics of {Brownian} motion and {Johnson} noise,''
  {\em American Journal of Physics}, vol.~64, no.~3, pp.~225--240, 1996.

\bibitem{Pascual-WinterPRB2012}
M.~F. Pascual-Winter, R.-C. Tongning, T.~Chaneli\`ere, and J.-L. Le~Gou\"et,
  ``Spin coherence lifetime extension in {Tm}${}^{3+}$:{YAG} through dynamical
  decoupling,'' {\em Phys. Rev. B}, vol.~86, p.~184301, Nov 2012.

\bibitem{Bar-GillNatComm2013}
N.~Bar-Gill, L.~M. Pham, A.~Jarmola, D.~Budker, and R.~L. Walsworth,
  ``{Solid-state electronic spin coherence time approaching one second},'' {\em
  Nature Communications}, vol.~4, pp.~1--6, apr 2013.

\bibitem{Genov2019MDD}
G.~T. Genov, N.~Aharon, F.~Jelezko, and A.~Retzker, ``Mixed dynamical
  decoupling,'' {\em Quantum Science and Technology}, vol.~4, p.~035010, jul
  2019.

\bibitem{Genov2020PRR}
G.~T. Genov, Y.~Ben-Shalom, F.~Jelezko, A.~Retzker, and N.~Bar-Gill,
  ``Efficient and robust signal sensing by sequences of adiabatic chirped
  pulses,'' {\em Phys. Rev. Research}, vol.~2, p.~033216, Aug 2020.

\bibitem{Aharon2016NJP}
N.~Aharon, I.~Cohen, F.~Jelezko, and A.~Retzker, ``Fully robust qubit in atomic
  and molecular three-level systems,'' {\em New Journal of Physics}, vol.~18,
  p.~123012, dec 2016.

\bibitem{barry2023sensitiveacdcmagnetometry}
J.~F. Barry, M.~H. Steinecker, S.~T. Alsid, J.~Majumder, L.~M. Pham, M.~F.
  O'Keeffe, and D.~A. Braje, ``Sensitive ac and dc magnetometry with
  nitrogen-vacancy-center ensembles in diamond,'' {\em Phys. Rev. Appl.},
  vol.~22, p.~044069, Oct 2024.

\bibitem{maze2012free}
J.~R. Maze, A.~Dr{\'e}au, V.~Waselowski, H.~Duarte, J.-F. Roch, and V.~Jacques,
  ``Free induction decay of single spins in diamond,'' {\em New Journal of
  Physics}, vol.~14, no.~10, p.~103041, 2012.

\bibitem{d2021introduction}
D.~d’Alessandro, {\em Introduction to quantum control and dynamics}.
\newblock Chapman and hall/CRC, 2021.

\bibitem{muller2022one}
M.~M. M{\"u}ller, R.~S. Said, F.~Jelezko, T.~Calarco, and S.~Montangero, ``One
  decade of quantum optimal control in the chopped random basis,'' {\em Reports
  on progress in physics}, vol.~85, no.~7, p.~076001, 2022.

\bibitem{vetter2024gate}
P.~J. Vetter, T.~Reisser, M.~G. Hirsch, T.~Calarco, F.~Motzoi, F.~Jelezko, and
  M.~M. M{\"u}ller, ``Gate-set evaluation metrics for closed-loop optimal
  control on nitrogen-vacancy center ensembles in diamond,'' {\em npj Quantum
  Information}, vol.~10, no.~1, p.~96, 2024.

\bibitem{hahn1950spin}
E.~L. Hahn, ``Spin echoes,'' {\em Physical review}, vol.~80, no.~4, p.~580,
  1950.

\bibitem{viola1998dynamical}
L.~Viola and S.~Lloyd, ``Dynamical suppression of decoherence in two-state
  quantum systems,'' {\em Phys. Rev. A}, vol.~58, pp.~2733--2744, Oct 1998.

\bibitem{meiboom1958modified}
S.~Meiboom and D.~Gill, ``Modified spin-echo method for measuring nuclear
  relaxation times,'' {\em Review of scientific instruments}, vol.~29, no.~8,
  pp.~688--691, 1958.

\bibitem{gullion1990new}
T.~Gullion, D.~B. Baker, and M.~S. Conradi, ``New, compensated carr-purcell
  sequences,'' {\em Journal of Magnetic Resonance (1969)}, vol.~89, no.~3,
  pp.~479--484, 1990.

\bibitem{uhrig2007keeping}
G.~S. Uhrig, ``Keeping a quantum bit alive by optimized
  $\ensuremath{\pi}$-pulse sequences,'' {\em Phys. Rev. Lett.}, vol.~98,
  p.~100504, Mar 2007.

\bibitem{ryan2010robust}
C.~A. Ryan, J.~S. Hodges, and D.~G. Cory, ``Robust decoupling techniques to
  extend quantum coherence in diamond,'' {\em Phys. Rev. Lett.}, vol.~105,
  p.~200402, Nov 2010.

\bibitem{souza2011robust}
A.~M. Souza, G.~A. \'Alvarez, and D.~Suter, ``Robust dynamical decoupling for
  quantum computing and quantum memory,'' {\em Phys. Rev. Lett.}, vol.~106,
  p.~240501, Jun 2011.

\bibitem{Souza2012}
A.~M. Souza, G.~A. Álvarez, and D.~Suter, ``Robust dynamical decoupling,''
  {\em Philosophical Transactions of the Royal Society A: Mathematical,
  Physical and Engineering Sciences}, vol.~370, pp.~4748--4769, 10 2012.

\bibitem{Genov2017PRL}
G.~T. Genov, D.~Schraft, N.~V. Vitanov, and T.~Halfmann, ``{Arbitrarily
  Accurate Pulse Sequences for Robust Dynamical Decoupling},'' {\em Physical
  Review Letters}, vol.~118, p.~133202, Mar 2017.

\bibitem{Ezzell23}
N.~Ezzell, B.~Pokharel, L.~Tewala, G.~Quiroz, and D.~A. Lidar, ``Dynamical
  decoupling for superconducting qubits: A performance survey,'' {\em Phys.
  Rev. Appl.}, vol.~20, p.~064027, Dec 2023.

\bibitem{rossignolo_quocs_2023}
M.~Rossignolo, T.~Reisser, A.~Marshall, P.~Rembold, A.~Pagano, P.~J. Vetter,
  R.~S. Said, M.~M. Müller, F.~Motzoi, T.~Calarco, F.~Jelezko, and
  S.~Montangero, ``{QuOCS}: The quantum optimal control suite,'' {\em Computer
  Physics Communications}, vol.~291, p.~108782, 2023.

\bibitem{zhang2015experimental}
J.~Zhang and D.~Suter, ``Experimental protection of two-qubit quantum gates
  against environmental noise by dynamical decoupling,'' {\em Phys. Rev.
  Lett.}, vol.~115, p.~110502, Sep 2015.

\bibitem{vallabhapurapu2022indirect}
H.~H. Vallabhapurapu, C.~Adambukulam, A.~Saraiva, and A.~Laucht, ``Indirect
  control of the ${}^{29}{\mathrm{siv}}^{\ensuremath{-}}$ nuclear spin in
  diamond,'' {\em Phys. Rev. B}, vol.~105, p.~205435, May 2022.

\bibitem{zhang2019improved}
J.~Zhang, S.~S. Hegde, and D.~Suter, ``Improved indirect control of nuclear
  spins in diamond {N}-{V} centers,'' {\em Phys. Rev. Appl.}, vol.~12,
  p.~064047, Dec 2019.

\bibitem{hegde2020efficient}
S.~S. Hegde, J.~Zhang, and D.~Suter, ``Efficient quantum gates for individual
  nuclear spin qubits by indirect control,'' {\em Phys. Rev. Lett.}, vol.~124,
  p.~220501, Jun 2020.

\bibitem{zhang2023coupling}
F.~Zhang, J.~Xing, X.~Hu, X.~Pan, and G.~Long, ``Coupling-selective quantum
  optimal control in weak-coupling {NV}-$^{13}${C} system,'' {\em AAPPS
  Bulletin}, vol.~33, no.~1, p.~2, 2023.

\bibitem{dolde2014high}
F.~Dolde, V.~Bergholm, Y.~Wang, I.~Jakobi, B.~Naydenov, S.~Pezzagna, J.~Meijer,
  F.~Jelezko, P.~Neumann, T.~Schulte-Herbr{\"u}ggen, {\em et~al.},
  ``High-fidelity spin entanglement using optimal control,'' {\em Nature
  communications}, vol.~5, no.~1, p.~3371, 2014.

\bibitem{Michaels2021multidimensional}
C.~P. Michaels, J.~Arjona~Mart{\'{i}}nez, R.~Debroux, R.~A. Parker, A.~M.
  Stramma, L.~I. Huber, C.~M. Purser, M.~Atat{\"{u}}re, and D.~A. Gangloff,
  ``Multidimensional cluster states using a single spin-photon interface
  coupled strongly to an intrinsic nuclear register,'' {\em {Quantum}}, vol.~5,
  p.~565, Oct. 2021.

\bibitem{bradley2019ten}
C.~E. Bradley, J.~Randall, M.~H. Abobeih, R.~C. Berrevoets, M.~J. Degen, M.~A.
  Bakker, M.~Markham, D.~J. Twitchen, and T.~H. Taminiau, ``A ten-qubit
  solid-state spin register with quantum memory up to one minute,'' {\em Phys.
  Rev. X}, vol.~9, p.~031045, Sep 2019.

\bibitem{van2024mapping}
G.~Van~de Stolpe, D.~Kwiatkowski, C.~Bradley, J.~Randall, M.~Abobeih,
  S.~Breitweiser, L.~Bassett, M.~Markham, D.~Twitchen, and T.~Taminiau,
  ``Mapping a 50-spin-qubit network through correlated sensing,'' {\em Nature
  Communications}, vol.~15, no.~1, p.~2006, 2024.

\bibitem{zahedian24blueprint}
M.~Zahedian, V.~Vorobyov, and J.~Wrachtrup, ``Blueprint for efficient nuclear
  spin characterization with color centers,'' {\em Phys. Rev. B}, vol.~109,
  p.~214111, Jun 2024.

\bibitem{joas2024high}
T.~Joas, F.~Ferlemann, R.~Sailer, P.~J. Vetter, J.~Zhang, R.~S. Said,
  T.~Teraji, S.~Onoda, T.~Calarco, G.~Genov, {\em et~al.}, ``High-fidelity
  electron spin gates in a scalable diamond quantum register,'' {\em arXiv
  preprint arXiv:2406.04199}, 2024.


\end{thebibliography}

\pagebreak
\renewcommand{\thefigure}{S.\arabic{figure}}
\setcounter{figure}{0}

\onecolumngrid
\pagebreak
\section*{Supplemental Material to ``Coherent Control of a Long-Lived Nuclear Memory Spin in a Germanium-Vacancy Multi-Qubit Node''}
\twocolumngrid

\section{Sample and setup description}
We work with a $\langle111\rangle$-oriented diamond, that was grown under high pressure high temperature (HPHT) conditions, during which Ge is naturally incorporated and formed GeV centers without additional annealing~\cite{palyanov2015germanium}. 
The investigated \gev center is located in a solid immersion lens (SIL) of 10\,\textmu m diameter and microwave (MW) and radio frequency (RF) fields are applied through a 20\,\textmu m-thick copper wire spanned over the diamonds surface nearby the SIL.
\\
A detailed setup description can be found in the supplementary of \cite{senkalla2024germanium}. 
We state here most important details and changes compared to previous configuration.
Experiments are performed in a dilution refrigerator (Bluefors, BF-LD400) with optical access for a home-built confocal microscope.
A superconducting vector magnet from American Magnetics, Inc. allows to precisely align the magnetic field with the \gev by optimizing electron spin lifetime at 4\,K, yielding $T_{1,e}^{\text{4K}}=(3.41\pm0.20)\,\text{ms}$.
As we observe a small decay of the magnetic field ($\sim$4.7\,\textmu T/h) in the persistent mode, we operate the magnet actively stabilized with magnetic field fluctuations of $std = 0.05\,\text{G}$.
For resonant excitation we use a Sirah Lasertechnik Matisse 2 DS single frequency tunable dye laser at 602\,nm (Rhodamin 6G in ethylene glycol).
Pulsed optical excitation is realized by using a double-path acousto optical modulator (AOM) configuration for which the distorted optical mode is cleaned by a single mode polarization maintaining patch cord fiber PM-S405-XP from Thorlabs. 
For in- and outcoupling we use a mounted aspheric lens form Thorlabs (A260TM-A).
After the outcoupler we use a 605/20 clean up filter and adjust the beam size to fill the back aperture of the Newport objective (Li-60X with NA of 0.85).
For MW and RF manipulation we use the positive and negative output of the same channel of the AWG (Tektronix 70000B HP).
The designated MW channel is filtered by a high- and low pass filter and pulses are amplified by an amplifier (ar 50S1G6).
Similarly, the designated RF channel is filtered by a DC block and low pass filter before the pulses are amplified by a second amplifier (ar 125A400).
The MW and RF lines are then combined by a diplexer (Procom PRO-DIPX225/330) and provided to the cryostat.
\section{Electron spin initialization and lifetime}
\begin{figure}
	\centering
	\includegraphics[width=8.6 cm]{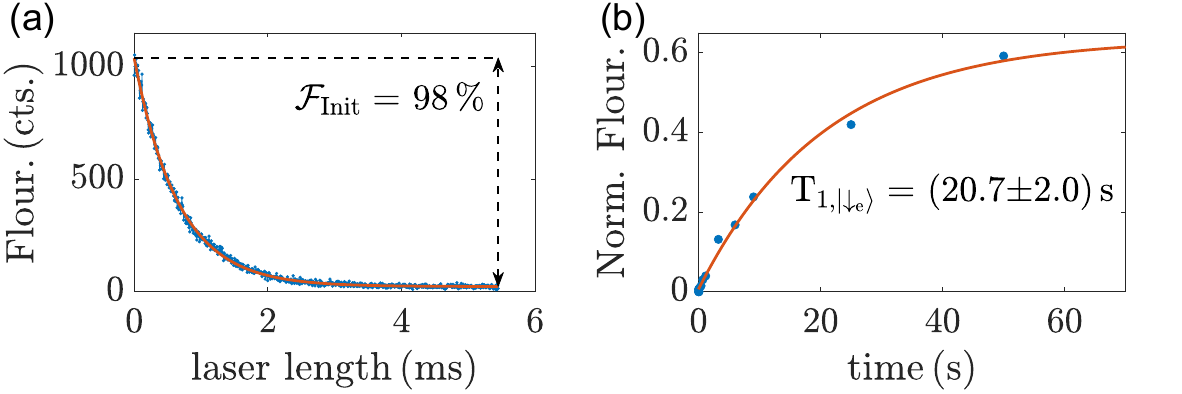}
	\caption{Electron spin properties.\quad(a) Exemplary initialization curve for the $\ket{\uparrow_{e}}$ state using a 5.5\,ms laser at optical transition $c_1$. (b) $T_{1,e}$ measurement in the $\ket{\downarrow_e}$ state.}
	\label{fig:supp_fig1}
\end{figure}

Figure \ref{fig:supp_fig1}(a) shows the fluorescence time trace of a 5.5\,ms long laser pulse resonant to $c_1$, applied to an initially prepared $\ket{\downarrow_e}$ state, from which we determine an initialization fidelity of 98$\,\%$.
The fidelity is calculated according to $1-\frac{a}{h}$, where $h$ and $a$ are values of the exponential fit at 0\,ms and 5.5\,ms, respectively. 
For the electron spin lifetime we measure $T_{1,e} = (20.7\pm2.0)\,\text{s}$, as shown in Fig.~\ref{fig:supp_fig1}~(b). 

\section{Hyperfine coupling parameter estimation}
\label{sec:supp_hyperfine}

\begin{figure}
	\centering
	\includegraphics[width=8.6 cm]{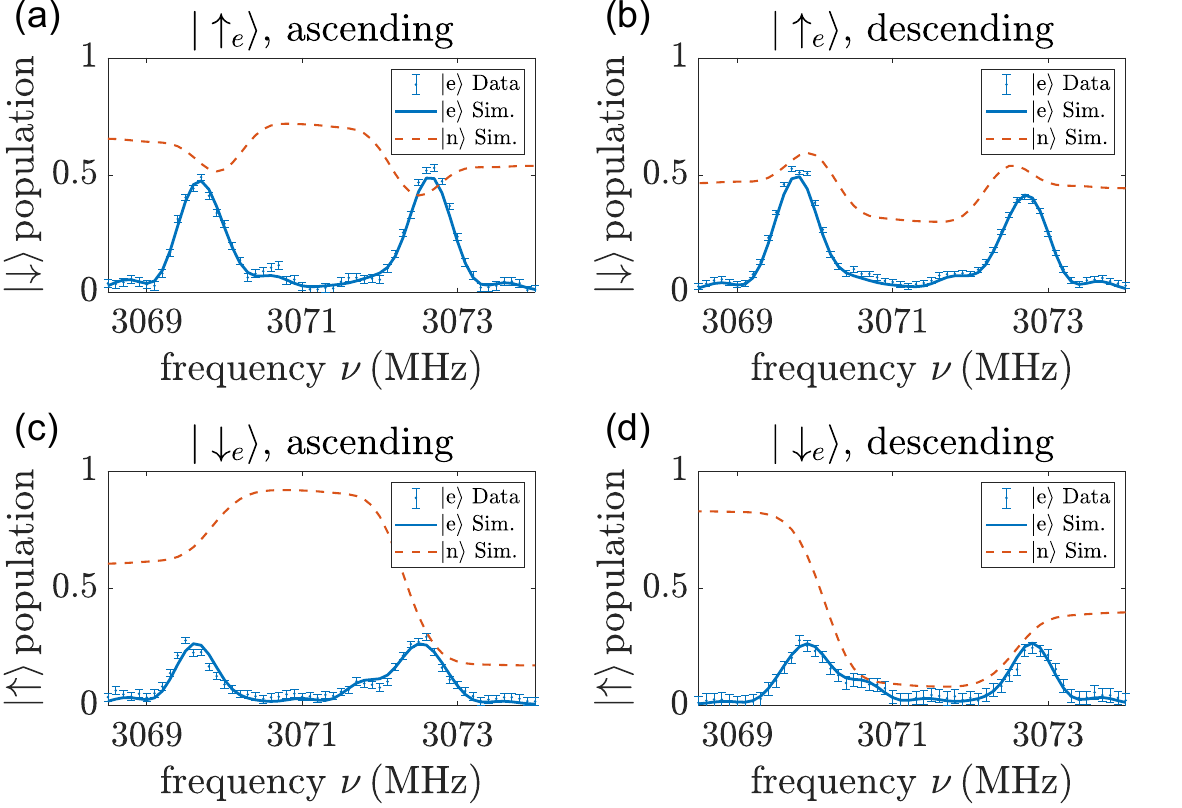}
	\caption{Pulsed ODMR measurements with simulation results. The fit is performed for four different experiments: (a) electron spin $\ket{\uparrow_e}$ with ascending frequency sweep direction, (b) electron spin $\ket{\uparrow_e}$ with descending frequency sweep direction, (c) electron spin $\ket{\downarrow_e}$ with ascending frequency sweep direction and (d) electron spin $\ket{\downarrow_e}$ with descending frequency sweep direction. The orange dotted line shows the nuclear spin population before the microwave pulse.}
	\label{fig:supp_fig2}
\end{figure}
We perform four pulsed ODMR measurements, which are shown in Fig.~\ref{fig:supp_fig2}\,(a)\,-\,(d) with their corresponding fit.
The four spectra differ by the frequency sweep direction and to which state the electron spin is re-initialized via the laser pulse.
Each spectrum is normalized by a Rabi experiment that is performed after each run of the sequence (see Fig. \ref{fig:supp_coherence}\,(a) for details).
The obtained measurement data corresponds to the average fluorescence signal of around 3470 repetitions of the measurement sequence (pulsed ODMR + Rabi).\\
To extract the hyperfine coupling parameters $A_{zx}$ and $A_{zz}$ between the electron and nuclear spin, we numerically simulate the obtained spectra according to the Hamiltonian Eq.~(1) from the main text. \\
The initial nuclear spin state $\rho^n_i= p_i\ket{\uparrow_n}\bra{\uparrow_n} + \left(1-p_i\right)\ket{\downarrow_n}\bra{\downarrow_n}$ is left open as a fit parameter since the measured spectra reflect the averaged fluorescence signal, and the nuclear spin can become polarized through the ODMR measurements and subsequent Rabi experiment.
The states $\ket{\uparrow_n}$ and $\ket{\downarrow_n}$ refer to the nuclear spin states for the diagonalized Hamiltonian along the corresponding quantization axis for a given electron spin state.
The nuclear spin is then initalized along the corresponding quantization axis. \\
Considering the low temperature and the resulting long nuclear spin population lifetimes, we model the laser pulse as a Kraus operator of a depolarizing channel~\cite{nielsen2010quantum}, fully re-initializing the electron spin to either $\ket{\uparrow_e}$ or $\ket{\downarrow_e}$, while leaving the nuclear spin untouched.
To account for the laser length of 5.5\,ms and the averaging over thousands of repetitions, we let the nuclear spin fully dephase through a randomized detuning. \\
Given the custom sample holder, the diamond is unlikely to be perfectly placed in the center of the magnetic field generated by the vector magnet.
The objective and other hardware components in the sample chamber may also cause slight deviation of the set magnetic field amplitude. 
To account for those effects, the strength of the magnetic field $B_i$ is treated as a fit parameter that can differ for each of the four spectra.
$B_i$ is then used to construct the hyperfine coupling parameters $A_{zx}$ and $A_{zz}$ according to Eq.~(2) and Eq.~(3) from the main text.
The nuclear transition frequencies $\omega_{\text{RF2}}=(2\pi)\,(493.62\pm0.04)\,\text{kHz}$ and $\omega_{\text{RF1}}=(2\pi)\,(2489.73\pm0.06)\,\text{kHz}$, required for this calculation, are extracted from the nuclear Ramsey fringes shown in Fig.~\ref{fig:supp_coherence}\,(b) and in Fig.~2\,(d) in the main text. \\
The detuning of Eq.~(1) in the main text is given by $\Delta_{i,k}=\gamma_e^\text{eff}B_i-(2\pi)\nu_k$, where the subscript $k$ is an index for the current measurement step, and $i=(a),(b),(c),(d)$ denotes that the magnetic field $B_i$ might vary in each experiment in Fig. \ref{fig:supp_fig2}(a,b,c,d).
Thus the detuning depends on both, the frequency step $\nu_k$ and the magnetic field $B_i$ from the particular ODMR sequence.
While the magnetic field $B_i$ varies between the four different ODMR spectra but is kept during one sequence, the gyromagnetic ratio $\gamma_e^\text{eff}$ is a shared property among all spectra.
Hereby, $\gamma_e^\text{eff}$ denotes the effective gyromagnetic ratio, which includes both, the electron Zeeman effect and the quenched orbital Zeeman effect.
While the gyromagnetic ratio of a free electron spin is well known, the orbital contribution of the \gev has not been experimentally determined yet~\cite{thiering2018ab}.
For this reason we combine both in the fit parameter $\gamma_e^\text{eff}$.
\\
We include the electron spin dephasing ($T_{2,e}^*=1.43\text{\,\textmu s}$~\cite{senkalla2024germanium}) during the microwave $\pi$-pulse by a randomized constant detuning with a standard deviation of $\sigma_\delta\approx\sqrt{2}/T_{2,e}^*=2\pi\cdot 146~\text{kHz}$~\cite{senkalla2024germanium} and a mean of 0.
Each simulated spectrum is then averaged over 2500 iterations.\\
We use differential evolution~\cite{storn1997differential} to find the set of parameters, i.e. the gyromagnetic ratio $\gamma_e^\text{eff}$, the magnetic field $B_i$, and the initial nuclear spin population $p_i$, that minimizes the sum of the squared differences between the measured and simulated spectra.
These parameters are then used as starting points in a final fit via the least-square method, that is individually performed for the four ODMR measurements while keeping the product $\gamma_e^\text{eff} B_i$ constant.
Our simulation fits the data very well.
We obtain an $R^2$ of 0.9854 for $\ket{\uparrow_e}$ and 0.9328 for $\ket{\downarrow_e}$ with an ascending frequency sweep direction and 0.9854 for $\ket{\uparrow_e}$ and 0.9568 for $\ket{\downarrow_e}$ with a descending frequency sweep direction.
This leads to the following parameters, where the subscript describes the electron spin manifold and the superscript the ODMR sweep direction:
\begin{itemize}
	\item $\gamma_{\ket{\uparrow_e}}^{ascending}=(2\pi)\,31.6162165179458\,\frac{\text{GHz}}{\text{T}}$,
	\item $\gamma_{\ket{\downarrow_e}}^{ascending}=(2\pi)\,31.6165307831433\,\frac{\text{GHz}}{\text{T}}$,
	\item $\gamma_{\ket{\uparrow_e}}^{descending}=(2\pi)\,31.6135603066038\,\frac{\text{GHz}}{\text{T}}$,
	\item $\gamma_{\ket{\downarrow_e}}^{descending}=(2\pi)\,31.6147971633278\,\frac{\text{GHz}}{\text{T}}$,
	\item $B_{\ket{\uparrow_e}}^{ascending}=97.140\text{\,mT}\pm 0.004\text{\,mT}$,
	\item $B_{\ket{\downarrow_e}}^{ascending}=97.138\text{\,mT}\pm 0.006\text{\,mT}$,
	\item $B_{\ket{\uparrow_e}}^{descending}=97.149\text{\,mT}\pm 0.004\text{\,mT}$,
	\item $B_{\ket{\downarrow_e}}^{descending}=97.146\text{\,mT}\pm 0.005\text{\,mT}$,
	\item $p_{\ket{\uparrow_e}}^{ascending}=0.34508 \pm 0.00004$,
	\item $p_{\ket{\downarrow_e}}^{ascending}=0.60563 \pm 0.00007$,
	\item $p_{\ket{\uparrow_e}}^{descending}=0.55639 \pm 0.00005$,
	\item $p_{\ket{\downarrow_e}}^{descending}=0.39752 \pm 0.00007$.
\end{itemize} 
According to Eq.~(2) and Eq.~(3) from the main text we then obtain the mean values $A_{zx}=2\pi\left(602.81\pm0.27\right)\text{\,kHz}$ and $A_{zz}=2\pi\left(2862.34\pm0.14\right)\text{\,kHz}$ for the hyperfine coupling parameters.
%

\subsection{Further error analysis}

\begin{figure}
	\centering
	\includegraphics[width=8.6 cm]{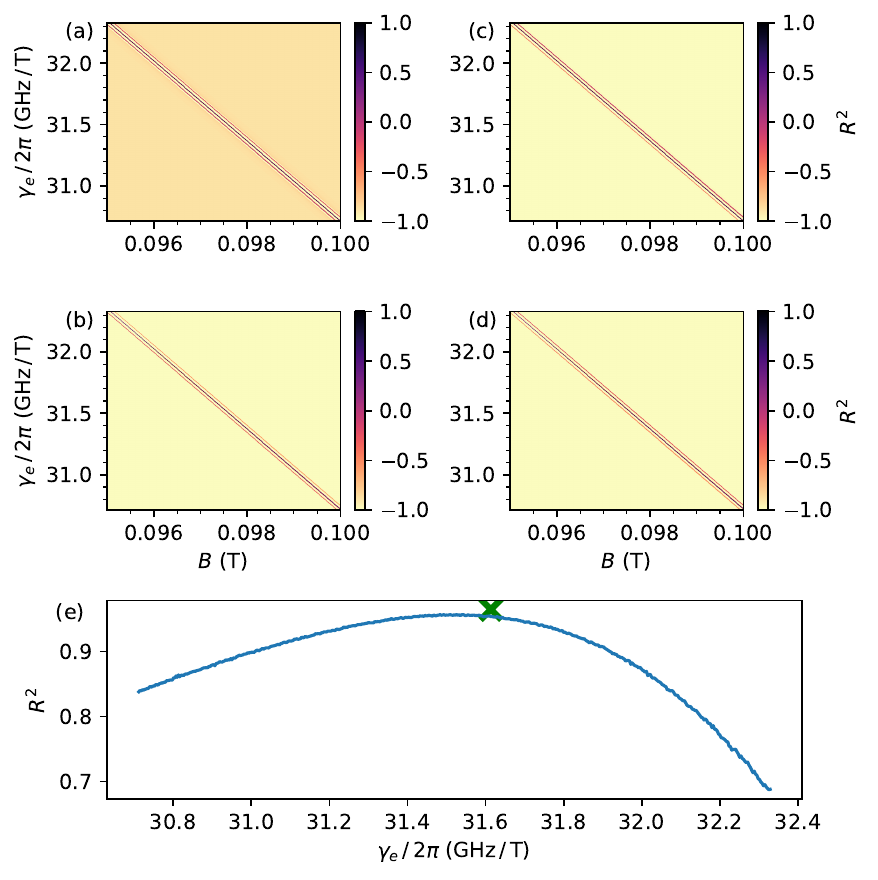}
	\caption{$R^2$ of the model for different values of $\gamma_e^{\text{eff}}$ and $B_i$. The images differ by their electron spin state of $\ket{\uparrow_e}$ for (a) ascending frequency sweep direction, (b) for descending frequency sweep direction and $\ket{\downarrow_e}$ for (c) ascending frequency sweep direction and (d) for descending frequency sweep direction. (e) Mean $R^2$ of the highest individually achieved $R^2$ of all four ODMR measurements for the respective $\gamma_e^{\text{eff}}$. The cross indicates the best mean $R^2$ value achieved in the simulation which is slightly higher than the mean $R^2$ values in the sweep due to the finite sampling size of the sweep.}
	\label{fig:supp_error_estimation}
\end{figure}

While only the magnetic fields $B_i$ go into the calculation of the hyperfine coupling parameters, $\gamma_e^{\text{eff}}$ and $B_i$ are coupled through the detuning $\Delta_i=\gamma_e^\text{eff}B_i-(2\pi)\nu$.
For this reason we kept the product $\gamma_e^\text{eff} B_i$ fixed while performing the final fit after the differential evolution.\\
To better understand the parameter landscape and the associated fit error, we sweep $\gamma_e^{\text{eff}}$ and $B_i$ in the range of $\lbrack30.712\,\frac{\text{GHz}}{\text{T}}, 32.328\,\frac{\text{GHz}}{\text{T}}\rbrack$ and $\lbrack 95\text{\,mT},100\text{\,mT}\rbrack$ for each of the four ODMR measurements and calculate the resulting $R^2$ value, as shown in Fig.~\ref{fig:supp_error_estimation}\,(a)\,-\,(d).
For each of the four plots we select the highest $R^2$ for all $\gamma_e^{\text{eff}}$ and visualize the mean value in Fig.~\ref{fig:supp_error_estimation}\,(e).
We observe a smooth parameter landscape with a distinct maximum.
We can thus conclude that our model accurately represents the experiment, and our fit results are precise, as the fit avoids being trapped in local maxima.

\section{Nuclear spin pumping}
\label{sec:nuc_spin_pumping}
\begin{figure}
	\centering
	\includegraphics[width=8.6cm]{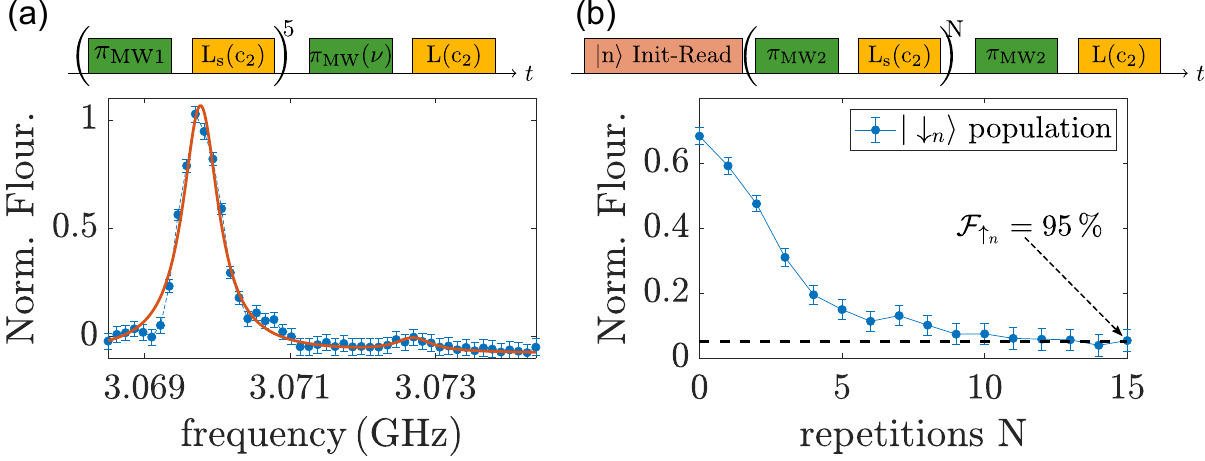}
	\caption{Nuclear spin initialization via spin pumping.\quad(a)~Nuclear spin pumping scheme using repetitive $\pi$ MW pulse and laser elements to initialize the nuclear spin probed with a pulsed ODMR. (b) Pulse sequence and experiment to estimate the initialization fidelity of the nuclear spin after $N$ repetitions using the spin pumping scheme. The population of the nuclear spin is prepared with the projection-SWAP gate.}
	\label{fig:rep_init_exp}
\end{figure}

\begin{figure}
	\centering
	\includegraphics[width=8.6cm]{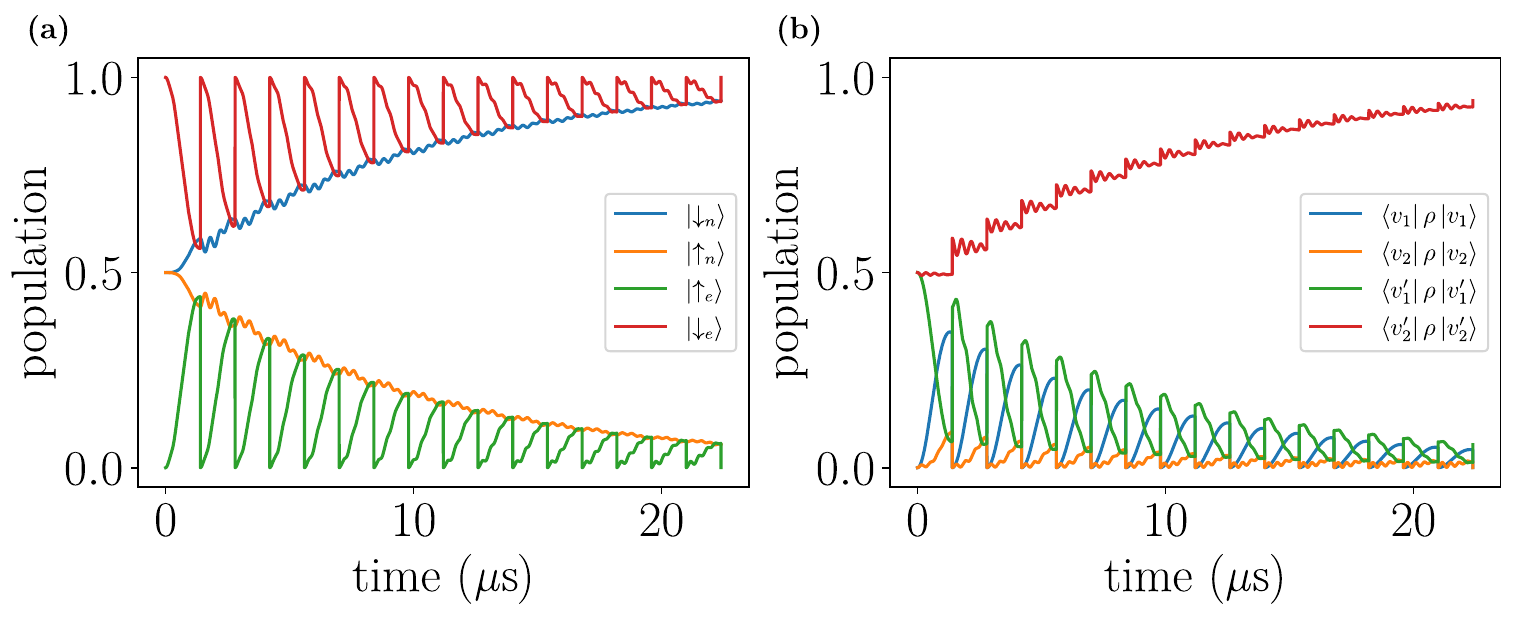}
	\caption{Nuclear spin initialization with electron resets (Simulation). The simulated fidelity for $N=15$ is 94\%, which is close to the experimental results. (a) The dynamics of the system if we trace out the other respective spin. (b) Dynamics in the eigenbasis within the $\ket{\downarrow_e}$ sub-manifold, with $v_i$ denoting the eigenvectors (see following section for definition).}
	\label{fig:rep_init_sim}
\end{figure}

Due to the strong hyperfine interaction, we already observe in the pulsed ODMR measurements that the nuclear spin can be polarized through the repetitive application of MW and laser pulses.
We can exploit this effect to implement a spin pumping scheme for initialization of the \nuc nuclear spin.
As illustrated in Fig.~1\,(b) in the main text, the nuclear spin experiences a different quantization axis depending on the state of the electron spin.
In our system, due to the large $A_{zx}$ and a $A_{zz}/2$ that is in the order of the bare nuclear Larmor frequency the angle between the quantization axes is significant ($\sim 30^{\circ}$).
As a result, the simultaneous flip of the electron and nucleus, which is typically forbidden, becomes partially allowed~\cite{EPR_jeschke}.
Now, to achieve polarization built-up, a spin pumping scheme has to meet two criteria.
First, considering the two hyperfine levels in an electron spin subspace, we need to be able to selectively depopulate one level while leaving the other untouched. 
In the remainder of this paragraph we will refer to the untouched level as dark state.
Due to the large energy gap $\omega_{\text{RF1}}$ this condition is well fulfilled in the $\ket{\downarrow_e}$ manifold, and choosing a narrow-band MW $\pi$-pulse at MW1 or MW2 can selectively depopulate one of the levels.
On the other hand, the small energy gap $\omega_{\text{RF2}}$ and the limited linewidth resolution, $\propto 1/(\pi T_{2,e}^*)$~\cite{Barry2020}, prohibits the formation of an efficient dark state in the $\ket{\uparrow_e}$ manifold. 
Second, we need a decay rate to the dark state.
For us, the laser leads to an electron reset causing a change of the quantization axis of the nuclear spin.
This creates a decay channel to the dark state while we pump the other state empty. 
\\
Hence, for the spin pumping scheme to work we start by preparing the electron spin in the $\ket{\downarrow_e}$ state with a 5.5\,ms laser pulse at $c_2$.
After application of a $\pi$-pulse on the MW1 or MW2 transition, we re-initialize the electron spin and repeat this process $N$ times.
In Fig.~\ref{fig:rep_init_exp}\,(a) the spin pumping scheme ($N=5$ repetitions) is probed with a pulsed ODMR experiment showing almost no population present in the $\ket{\uparrow_n}$ state. 
However, due to the change in nuclear spin polarization throughout the ODMR measurement, we cannot determine the initialization fidelity with the ODMR measurement after a specific number of repetitions $N$.
Thus, we equally prepare the initial nuclear spin state for every point with the projection-SWAP in Fig.~\ref{fig:rep_init_exp}\,(b).
This allows us to estimate an initialization fidelity of the nuclear spin using the spin pumping scheme of $95\,\%$ after repeating the sequence $N=15$ times (MW $\pi$ pulse duration 1.4$\,\mu$s).
Numeric simulations displayed in Fig.~\ref{fig:rep_init_sim}, accounting for electron spin dephasing, show that we can reach up to $98.7\,\%$ polarization of the nuclear spin by repeating this 50 times,
with the same $\pi$ MW pulse duration of $\SI{1.4}{\micro \s}$ as in the experiment.

\section{Driving field optimization with QuOCS}\label{sec:quocs}
\begin{figure}
	\centering
	\includegraphics[width=8.6 cm]{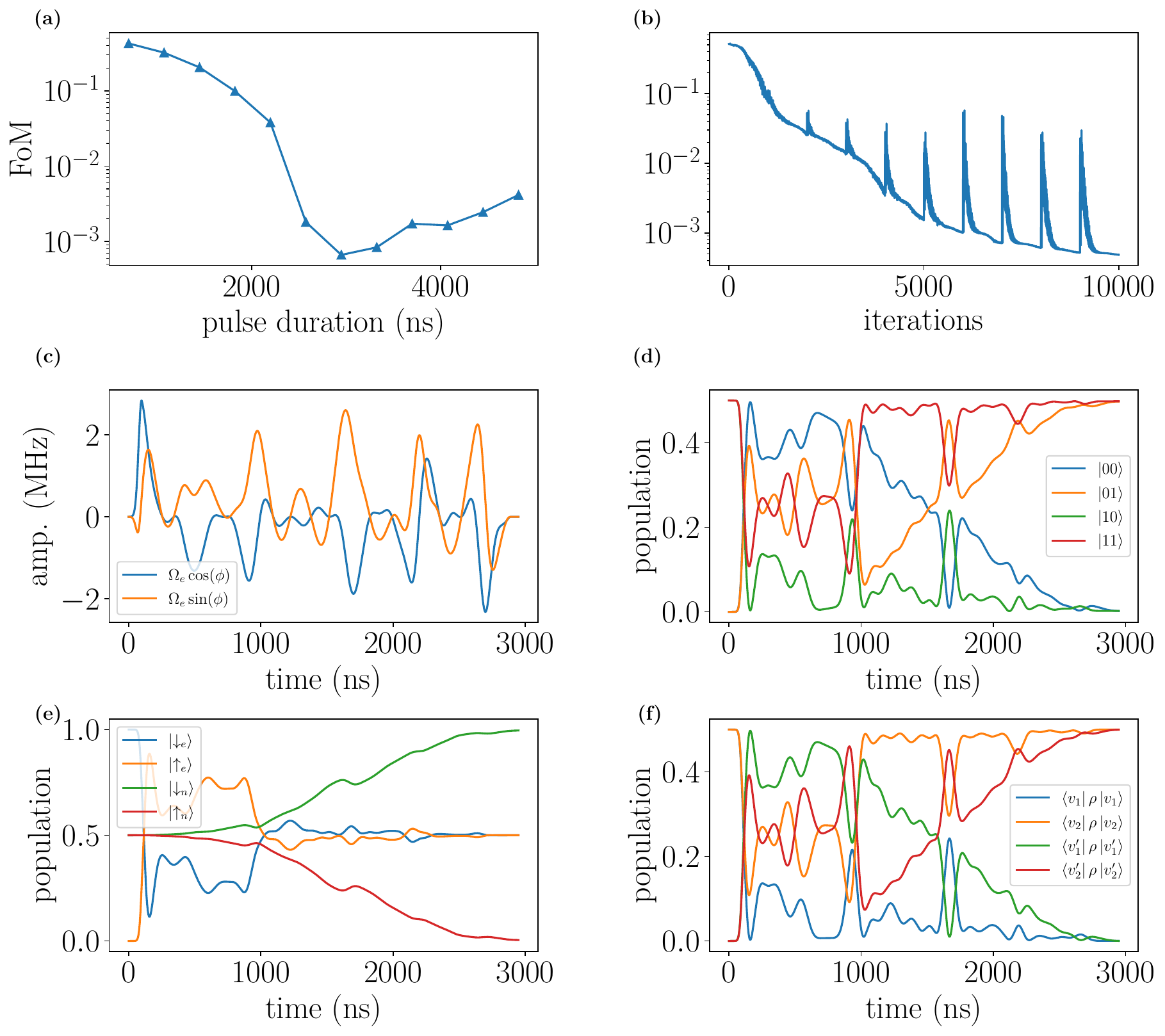}
	\caption{Optimization process for the initialization of the nuclear spin with a single optimized MW pulse. (a) Resulting polarization infidelity of the nucleus for different pulse durations, where each point was optimized separately (averaged over 5000 OU noise realizations). (b) Optimization progress for a fixed pulse duration (Optimization is done for 100 fixed OU noise realizations). (c) Resulting driving pulse shapes for one optimized duration. (d)-(f) Time evolution (d) computational basis, (e) tracing out the other spin, (f) populations along the nuclear quantization axis defined by $\ket{\downarrow_e}$ during the application of an optimized pulse. (b) - (f) relate to a pulse duration of $\SI{2950}{\nano \s}$.}
	\label{fig:supp_fig3}
\end{figure}

With the fitted Hamiltonian parameters, we can optimize the driving field to implement an operation of our choice.
The corresponding Hamiltonian (in the rotating frame at the driving field frequency after using the rotating wave approximation) can be written as
\begin{align}
	H_d = \Omega_e(t) \left[\cos\phi(t) S_x + \sin\phi(t) S_y \right]
\end{align}
where $\Omega_e(t)$ is the time-dependent Rabi frequency and $\phi(t)$ is its time-dependent phase. The driving frequency is chosen to be $\omega_d=\omega_\text{MW2}+\omega_\text{RF1}$.\\
With optimal control, we can now design functions which implement a population transfer procedure.
Because the nuclear spin eigenstate depends on the electron state, we choose to polarize along the quantization axis defined by the electron spin being in $\ket{\downarrow_e}$ 
\begin{align}
	\text{FoM} = 1 - \bra{v_2} \rho \ket{v_2} - \bra{v_2'} \rho \ket{v_2'}
\end{align}
where $v_2 (v_1)$ is the eigenvector corresponding to $\ket{\downarrow_e \uparrow_n} (\ket{\downarrow_e \downarrow_n})$. We introduced here $v_i' = (\sigma_x \otimes \mathbf{1}) v_i$, which point for the nucleus subspace in the same direction as the eigenvectors in the $\ket{\downarrow_e}$-subspace but have a flipped electron spin state.
We minimize the FoM using the QuOCS-Software \cite{rossignolo_quocs_2023} by optimizing $\Omega_e(t)$ and $\phi(t)$ with the dCRAB-algorithm~\cite{rach2015,muller2022one}. \\
We restrict our search space to functions in the Fourier-basis domain, limit the maximal Rabi-frequency and ensure with a flattop Gaussian envelope (raise time of 100\,ns) that we do not have sudden jumps in the driving amplitude.\\
We then start the optimization  (settings for dCRAB: 10 super iterations, 1000 function evaluations per super iteration) for different evolution times $\tau$.
The optimization process and its results are discussed in Figure \ref{fig:supp_fig3}.
According to the simulations, an efficient population transfer with high fidelity can be realized, achieving a minimum average nuclear polarization of $99.9\%$, without necessitating electron reinitialization.
%
\section{Further details on \nuc spin coherence experiments}
\label{sec:supp_additional}
As shown in the main text, we initialize and readout the nuclear spin by a projection-SWAP gate used for population transfer. 
Depending on the electron spin state we use a different MW and laser transition for the sequence in Fig. 2\,(a) of the main text. 
Here, the subscript \textit{s} (silenced) for a laser element indicates that we do not record data for this laser element.

\begin{figure}
	\centering
	\includegraphics[width=8.6 cm]{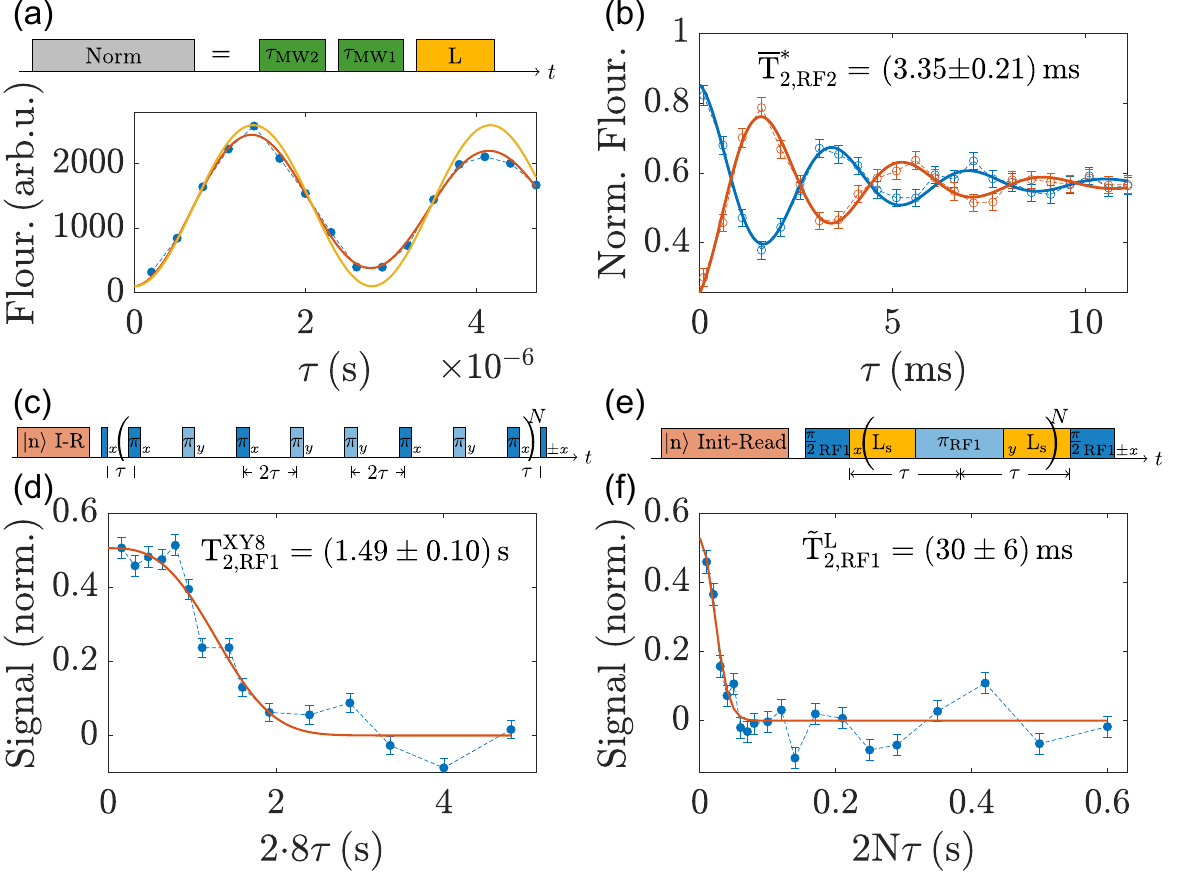}
	\caption{(a)~Top: Rabi sequence used for normalization that is appended after each measurement. Bottom: Exemplary normalization Rabi for the nuclear XY8 measurement shown in (d). Respective counts are normalized on the maximum and minimum of the non-decayed fit (yellow). (b)~Nuclear Ramsey in the $\vert\uparrow_{e}\rangle$ state showing a $T_{2,\text{RF2}}^*=(3.35\pm0.21)\text{\,ms}$. (c)~XY8 pulse sequence. (d)~Nuclear XY8 measurement in $\ket{\downarrow_e}$ yields a $T_{2,\text{RF1}}^{\text{XY8}}=(1.49\pm0.10)\,\text{s}$. (e)~CPMG sequence with laser applied during interpulse spacing $\tau$. (f)~Nuclear CPMG N=1 with the pulse sequence from (e). We observe reduced $T_2$ from $T_{2,\text{RF1}}^{\text{CPMG1}}=(270\pm11)\,\text{ms}$ to $\tilde{T}_{2,\text{RF1}}^{\text{L}}=(30\pm6)\,\text{ms}$.}
	\label{fig:supp_coherence}
\end{figure}

We initialize our two-qubit system in the respective eigenbasis of our diagonalized Hamiltonian.
For $\ket{\downarrow_e}$ we prepare the system in $\ket{\downarrow_e\downarrow_n}$ using a laser at $c_2$ and $\pi$-pulses at MW1 and RF2.
For $\ket{\uparrow_e}$ we initialize in $\ket{\uparrow_e\downarrow_n}$ using a laser at $c_1$ and $\pi$-pulses at MW2 and RF2.
The corresponding Rabi frequency used for RF2 is $\Omega_{\text{RF2}}=(2\pi)(7.73\pm0.05)\,\text{kHz}$ inferring a $\pi$ RF pulse length of 65\,\textmu s.
Furthermore, as outlined in the main text, each measurement is followed by an electron Rabi normalization sequence, which includes two successive conditional MW pulses, as depicted in Fig.~\ref{fig:supp_coherence}(a).
The maximum and minimum of the sinusodial part of the fit is used to normalize the obtained fluorescence of the respective measurement.\\
Fig. \ref{fig:supp_coherence}(b) shows $T_{2,\text{RF2}}^*=(3.35\pm0.21)$\,ms for the nuclear spin within the $\ket{\uparrow_e}$ manifold.
This longer $T_2^*$, compared to the one shown in the main text for $\ket{\downarrow_e}$, can be partly attributed to magnetic field fluctuations having a lower impact in the $\ket{\uparrow_e}$ state. 
This can be explained by noting that the contribution of the unperturbed term $(A_{zx}/2)^2$ to $\omega_{\text{RF1,RF2}}$ is greater in the $\ket{\uparrow_e}$ state than in the $\ket{\downarrow_e}$ state. Specifically, in the $\ket{\uparrow_e}$ state, $\omega_{\text{RF2}}^2$ is given by $(A_{zz}/2 - \gamma_n B)^2 + (A_{zx}/2)^2$, which includes a larger contribution from $(A_{zx}/2)^2$.
Further, the variations in the $T_2^*$ values of the \nuc could be attributed to slight differences in experimental conditions between the two measurements conducted during separate cooling cycles.
However, further investigation is required to fully explain this phenomena. \\
To investigate the influence of pulse errors on our nuclear spin coherence time we perform a XY8 dynamical decoupling sequence and compare it to the CPMG-8 one.
The pulse sequence and the resulting measurement for XY8 are shown in Fig.~\ref{fig:supp_coherence}\,(c) and (d), respectively.
We obtain $T_{2,\text{RF1}}^{\text{XY8}}=(1.49\pm0.10)\,\text{s}$ for XY8, which is comparable to the CPMG-8 result $T_{2,\text{RF1}}^{\text{CPMG8}}=(1.36\pm0.14)\,\text{s}$.
Additionally, to estimate the laser-induced decoherence on the \nuc nuclear spin, we examine a scenario with a continuous readout laser pulse at $c_2$ during the interpulse spacing $\tau$ of a CPMG-1 measurement in $\vert\downarrow_e\rangle$ shown in Fig.~\ref{fig:supp_coherence}\,(e).
We observe a reduce of the coherence time $T_2$ by a factor of 9 from $T_{2,\text{RF1}}^{\text{CPMG1}}=(270\pm11)\,\text{ms}$ to $\tilde{T}_{2,\text{RF1}}^{\text{L}}=(30\pm6)\,\text{ms}$ [Fig.~\ref{fig:supp_coherence}\,(f)]. 
While the readout laser $c_2$ is not in resonance with the electron spin manifold $\vert\downarrow_e\rangle$, residual overlap in the optical transition creates a small probability of exciting the electron spin, potentially leading to a spin flip.
Such an uncontrolled flip of the electron spin causes the \nuc nuclear spin to decohere, which is especially significant for our strong hyperfine interaction.
Recent studies on hyperfine interactions in the excited state \cite{beukers2024control} indicate that any phase acquired is linear with time spent in the excited state, which can be compensated already by simple decoupling sequences as the Hahn echo or CPMG N=1.
However, due to our strong hyperfine coupling it is unclear whether this phase remains linear for our system.
\\
Assuming e.g.~that the electron spin is in the state off-resonant to the laser pulse, with a readout window of $660\,\mathrm{\mu s}$ (1/e of laser pulse duration) we would be able to readout the electron spin 45 times before the nuclear spin loses its coherence.
\section{Numerical simulation and estimation of the memory time limit}

\begin{figure}
	\centering
	\includegraphics[width=8 cm]{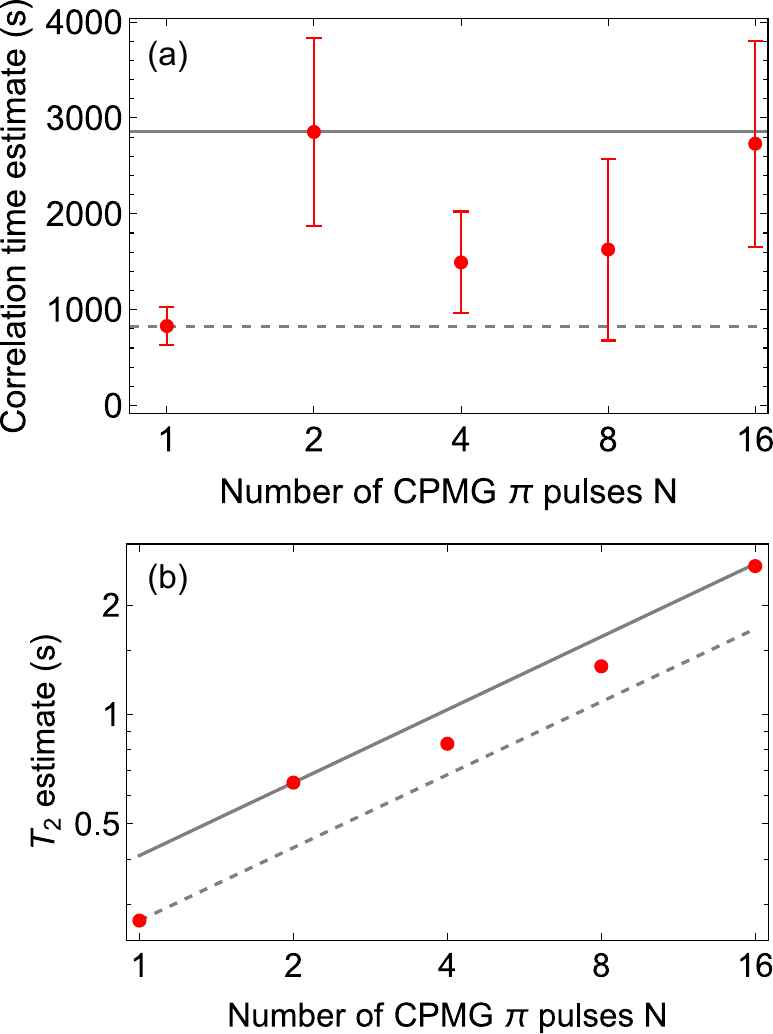}
	\caption{(a) Estimated correlation time $\tau_c$ of the OU model decay rate for different CPMG sequences vs. the number $N$ of the $\pi$ pulses in the sequence with $N=1$, corresponding to the standard Hahn echo (see also Fig. 3 in the main text).  
		(b) Experimentally measured $T_2$ times of the CPMG sequences vs. the number $N$ of the $\pi$ pulses in the sequence. The dashed (solid) lines are estimated with the formula in Eq. \eqref{Eq:OU_formula_approx_T2_tau} correspond to the expected minimum (maximum) values from the OU model, given the respective estimates for the correlation time $\tau_c$ in (a), showing the expected $N^{2/3}$ dependence, within the respective error range.  
	}
	\label{Fig:CPMG_tau_T2_estimation}
\end{figure}

In order to characterize the performance of our dynamical decoupling sequences in Fig.~3 in the main text for extending the coherence time of the $^{13}$C nuclear spin, we consider a simplified model of a two-state quantum system, which is subject to magnetic noise. In addition, we model the  power fluctuations of the driving fields and simulate the system evolution numerically. 
The model is similar to the one used for modeling decoherence of color centers in diamond 
\cite{senkalla2024germanium}. 
We consider the Hamiltonian
\begin{align}
	H(t)=&\left(\omega_0+\delta(t)\right)S_{z}\notag\\
	&+2\Omega_{\text{RF}} f(t)(1+\epsilon(t))\cos{\left[\omega_0 t+\phi(t)\right]}S_{x},
\end{align}
where $\omega_0$ is the expected Larmor frequency of the $^{13}$C nuclear spin, $\Omega_{\text{RF}}$ is the peak Rabi frequency of the radio-frequency (RF) driving field, $f(t)$ characterizes target variation of the Rabi frequency in time, e.g., it can be a step function taking values $0$ and $1$, while $\phi(t)$ is the phase of the applied field, $S_j=\hbar \sigma_i/2$ with $j\in \{x,y,z\}$ are the respective spin-$1/2$ operators. 
The parameters $\delta(t)$ and $\epsilon(t)$ characterize the time-varying errors in the Larmor frequency of the the $^{13}$C nuclear spin and the target Rabi frequency.\\
We move to the interaction picture with respect to $H_0^{(1)}=\omega_0 S_{z}$ and obtain after applying the rotating-wave approximation ($\Omega_{\text{RF}}\ll\omega_0$)
\begin{align}\label{H_sensing_1s}
	H_{1}(t)&=\delta(t)S_{z}\\
	&+\Omega_{\text{RF}}f(t)(1+\epsilon(t))(\cos{(\phi(t))}S_{x}+\sin{(\phi(t))}S_{y}).\notag
\end{align}
We consider this Hamiltonian in the simulation as the rotating-wave approximation is satisfied very well for our experimental parameters.\\
The noise of the detuning $\delta(t)$ is modelled as an Ornstein-Uhlenbeck (OU) process \cite{UhlenbeckRMP1945,Gillespie1996AJP} with a zero expectation value $\langle \delta(t)\rangle = 0$, correlation function $\langle \delta(t)\delta(t^{\prime})\rangle =\sigma_{\delta}^2\exp{(-|t-t^{\prime}|/\tau_c)}$, where $\sigma_{\delta}^2=\langle \delta(t)^2\rangle$ is the variance of the detuning due to noise, which is characterized by the standard deviation $\sigma_{\delta}=\sqrt{D\tau_c/2}$ with $D$ a diffusion constant, and $\tau_c$ is the correlation time of the noise for the OU process \cite{Gillespie1996AJP}. More details about the simulation implementation can be found in \cite{senkalla2024germanium}. \\
We calibrate the effect of environmental noise from the decay time $T_2^{\ast}\approx 2~$ms of the signal from a Ramsey measurement (see Fig. 2\,(d) in the main text), the decay time $T_2^H\approx\,274~$ms of a Hahn echo measurement, where a $\pi$ pulse is applied in the middle of the interaction (see Fig. 2\,(e) in the main text), and the decay times of the CPMG sequences in Fig. 3 in the main text. 
We note that our spin echo measurement consists of a single block with a duration $\widetilde{\tau}=2\tau$, which includes the refocusing $\pi$ pulse. In the following we use $\widetilde{\tau}=2\tau$ correspond to the typical pulse separation, used for the dynamical decoupling sequences we use. 
We then obtain the correlation time $\tau_c$ by fitting the exact analytical formula for the expected coherence decay rate $\gamma(N,\widetilde{\tau})$ due to magnetic noise (the signal $\sim \exp\{-\gamma(N,\widetilde{\tau})\}$), modelled with an OU process, during a sequence of $N$ ideal, instantaneous $\pi$ pulses for total evolution time $t=N \widetilde{\tau}=2N\tau$ \cite{senkalla2024germanium,Pascual-WinterPRB2012}:
\begin{widetext}
	\begin{equation}\label{Eq:OU_formula_full}
		\gamma(N,\widetilde{\tau})=\sigma _{\delta }^2\tau_c ^2  \left[-\left((-1)^{\text{N}+1} e^{-\frac{t}{\tau_c}}+1\right)
		\left(1-\text{sech}\left(\frac{\widetilde{\tau}}{2 \tau_c }\right)\right)^2+t
		\left(\frac{1}{\tau_c }-\frac{2 \tanh \left(\frac{\widetilde{\tau}}{2 \tau_c
			}\right)}{\widetilde{\tau}}\right)\right].
	\end{equation}
\end{widetext}
In the usual case when the noise correlation time is much longer than the pulse separation $\tau_c\gg \tau$, as is the case in our experiment, the decay rate simplifies to
\begin{equation}\label{Eq:OU_formula_approx}
	\gamma_{\text{approx}}(N,\widetilde{\tau})\approx \sigma_{\delta }^2  \frac{N\widetilde{\tau} ^3}{12\tau_c} .
\end{equation}
In an experiment where we fix the number of $\pi$ pulses $N$ and vary the pulse separation time $\widetilde{\tau}$ (as in the experiments in Fig. 3 in the main text), the decay rate as a function of the evolution time $t$ is given by  
\begin{equation}
	\gamma_{\text{approx}}(N,t)\approx \sigma_{\delta }^2  \frac{t^3}{12 N^2\tau_c},
\end{equation}
where we replaced $\widetilde{\tau}=t/N$ in the formula of Eq. \eqref{Eq:OU_formula_approx}. We determine the theoretical coherence times $T_2(N)$ for the different orders of the CPMG sequences by solving for $\gamma_{\text{approx}}(N,T_2)=1$, which leads to 
\begin{equation}\label{Eq:OU_formula_approx_T2_tau}
	T_{2}(N)\approx\left(\frac{12 N^2\tau_c}{\sigma _{\delta }^2} \right)^{1/3}=T^{H}_{2}N^{2/3},
\end{equation}
where $T^{H}_{2}=\left(\frac{12\tau_c}{\sigma _{\delta }^2} \right)^{1/3}$ the decay time of a Hahn echo experiment. The dependence of the coherence times $\sim N^{2/3}$ has been experimentally demonstrated in color centers in diamond \cite{Bar-GillNatComm2013}.\\
Figure \ref{Fig:CPMG_tau_T2_estimation}(a) shows the estimated correlation time for different CPMG sequences vs. the number $N$ of the $\pi$ pulses in the sequence. We fit the decay rate function in Eq. \eqref{Eq:OU_formula_full} to the experimental data, using NonLinearModelFit procedure in Mathematica. The correlation time estimates vary between $829\pm\,197$ s for $N=1$ (Hahn echo) and $2854\pm929$ s for ($N=2$). The estimates are within the error range, which is the 95\% confidence interval for all estimates, except for $N=1$. The slightly lower value for Hahn echo is most likely due to pulse errors, which are typically well compensated for the CPMG sequences for the particular initial state after the first $\pi/2$ pulse. 
Figure \ref{Fig:CPMG_tau_T2_estimation}(b) shows that the coherence times follow the expected $N^{2/3}$ behavior within the error range of the estimated correlation times. Finally, we note that in order to be conservative in our estimation of the theoretical estimation of the memory time limit in the main text (Fig. 3(d)), we use the lower bound of the correlation time estimate, i.e. $\tau_c=829$ s in all theoretical calculations and simulations. \\
Quantum memory experiments typically use fixed, optimized pulse separation time $\widetilde{\tau}$ and vary the number $N$ of the applied $\pi$ pulses. Then, the signal decay rate depends mainly on $\widetilde{\tau}$ and is given by
\begin{equation}
	\gamma_{\text{approx}}(\widetilde{\tau},t)\approx \sigma _{\delta }^2\frac{\widetilde{\tau} ^2 }{12 \tau_c}  t,
\end{equation}
where we replaced $N=t/\widetilde{\tau}$ in the formula of Eq. \eqref{Eq:OU_formula_approx}. We determine the theoretical memory times $T^{\text{mem}}_2(\widetilde{\tau})$ for the different pulse separation times of the CPMG sequences by solving for $\gamma_{\text{approx}}(\widetilde{\tau},T_2)=1$, which leads to 
\begin{equation}
	T^{\text{mem}}_2(\widetilde{\tau})\approx\frac{12 \tau_c}{\sigma _{\delta }^2\widetilde{\tau}^2}=\left(\frac{T_{2}^{H}}{\widetilde{\tau}}\right)^2 T_{2}^{H}.
\end{equation}
This is the formula we use in the main text to estimate the maximum achievable memory time, using dynamical decoupling with ideal, instantaneous $\pi$ pulses. It is evident that, assuming ideal, instantaneous pulses and no detrimental heating effect, reducing the pulse separation time $\widetilde{\tau}$ increases significantly the achievable memory time. We note that reducing the pulse separation time (increasing the number of $\pi$ pulses) is likely to reach a limit when we take pulse errors into account, so we analyze this next by simulating the process numerically. \\
Apart from the theoretical limit for the OU noise model, we also simulate the effect that the $\pi$ pulses are not instantaneous and there are also pulse errors due to detuning and amplitude noise. We described the characteristics of detuning noise above. We model the fluctuations in the driving field amplitude similarly to previous experiments as they typically depend on the used arbitrary waveform generator and amplifier \cite{Genov2019MDD,Genov2020PRR,senkalla2024germanium}. Specifically, we model the relative error of the Rabi frequency with an OU process with the update function \cite{Genov2019MDD,senkalla2024germanium}
\begin{equation}\label{Eq:OU_noise_epsilon}
	\epsilon(t+\Delta t)=\epsilon(t)e^{-\frac{\Delta t}{\tau_{\Omega}}}+\widetilde{n_{\epsilon}}\sqrt{\sigma_{\epsilon}^2\left(1-e^{-\frac{2\Delta t}{\tau_{\Omega}}}\right)},
\end{equation}
where $\sigma_{\epsilon}=\sqrt{(1/2)D_{\Omega}\tau_{\Omega}}=0.005$, the correlation time $\tau_{\Omega}=500 \mu$s with the corresponding diffusion constant $D_{\Omega}=2\sigma_{\epsilon}^2/\tau_{\Omega}$ \cite{Aharon2016NJP}.
The initial values of $\delta(t=0)$ and $\epsilon(t=0)$ change from run to run and are taken from a Gaussian distribution with standard deviations $\sigma_{\delta}$ and $\sigma_{\epsilon}$, respectively. The results are demonstrated in the inset of Fig. 3\,(b) in the main text. Theoretically, one expects a prolongation of the memory time with shorter pulse separation when the pulses are ideal and instantaneous, which is shown by the solid red line within the Figure. We show that when the pulses are not instantaneous, pulse errors become an important factor when the pulse separation is of the order of 20\,ms or shorter. The best possible memory time is then achieved at pulse separation $\widetilde{\tau}=2\tau\approx 10$\,ms. Longer pulse separation does not allow for efficient decoherence suppression for our current noise spectrum while shorter pulse spacing requires many pulses and the performance suffers from pulse errors. Our current experimental setup shows no noticeable heating effect for pulse separation of $24$\,ms or higher (c.f. subsequent section), so reaching pulse distance of $10$\,ms might be feasible with modest modifications.\\
Finally, the achievable memory times are also affected by $T_{1,e}$ relaxation. The memory time limit $T^{\text{mem}}_{2,\text{total}}$ that takes $T_{1,e}$ into account is then determined by the formula 
\begin{equation}
	\frac{1}{T^{\text{mem}}_{2,\text{total}}(\widetilde{\tau})}=\frac{1}{2T_{1,e}}+\frac{1}{T^{\text{mem}}_{2}(\widetilde{\tau})},
\end{equation}
which we use in Fig. 3\,(b) inset in the main text. We show that for pulse separation of the order of 40\,ms or shorter, the main memory decay factor is $T_{1,e}$ relaxation. We estimate a  memory time of $T^{\text{mem}}_2=18.1\,$s with pulse separation of 24\,ms, which is feasible in our experiment without excessive heating. Reducing pulse separation to around 10\,ms by improved MW and RF delivery should boost the memory time to $T^{\text{mem}}_2\approx 28\,$s where pulse errors with the XY8 sequence start to play a role due to the large number of pulses. Application of higher order sequences like KDD or the UR family~\cite{Souza2012,Genov2017PRL,Ezzell23} can in principle boost the memory time limit even closer to the $T_{1,e}$ limit of $\approx 41.4\,$s. 

In the following, we characterize the fidelities of the radio-frequency pulses, which are used in Fig. 3 in the main text. In order to do this we define their fidelity similarly to \cite{Souza2012,Genov2017PRL,senkalla2024germanium} as 
\begin{equation}
	F(t)=\frac{1}{2}\left|\text{Tr} \left(U_{1,\text{no noise}}^\dagger(t,0)U_{1}(t,0)\right)\right|,
\end{equation}
where $U_{1,\text{no noise}}(t,0)=\exp\left(-i H_1 t\right),\delta(t)=0,\epsilon(t)=0$ is the target propagator of the pulse without noise and $U_{1}(t,0)=\exp\left(-i H_1 t\right),\delta(t)\ne 0,\epsilon(t)\ne 0$ is the actual one. As the correlation time of the noise is much longer than the duration of the $\pi/2$ and $\pi$ pulses,  we assume for simplicity that the frequency and amplitude noise terms $\delta(t)$ and $\epsilon(t)$ are constant during the interaction. We note that, similarly to Fig. 3 in the main text, we also assume that the nuclear spin measurements are conducted in the $\ket{\downarrow_e}$ manifold, i.e., the electron spin is initialized and remains in state $\ket{\downarrow_e}$ during the experiments. 
%

\begin{figure*}
	\includegraphics[width=0.95\textwidth]{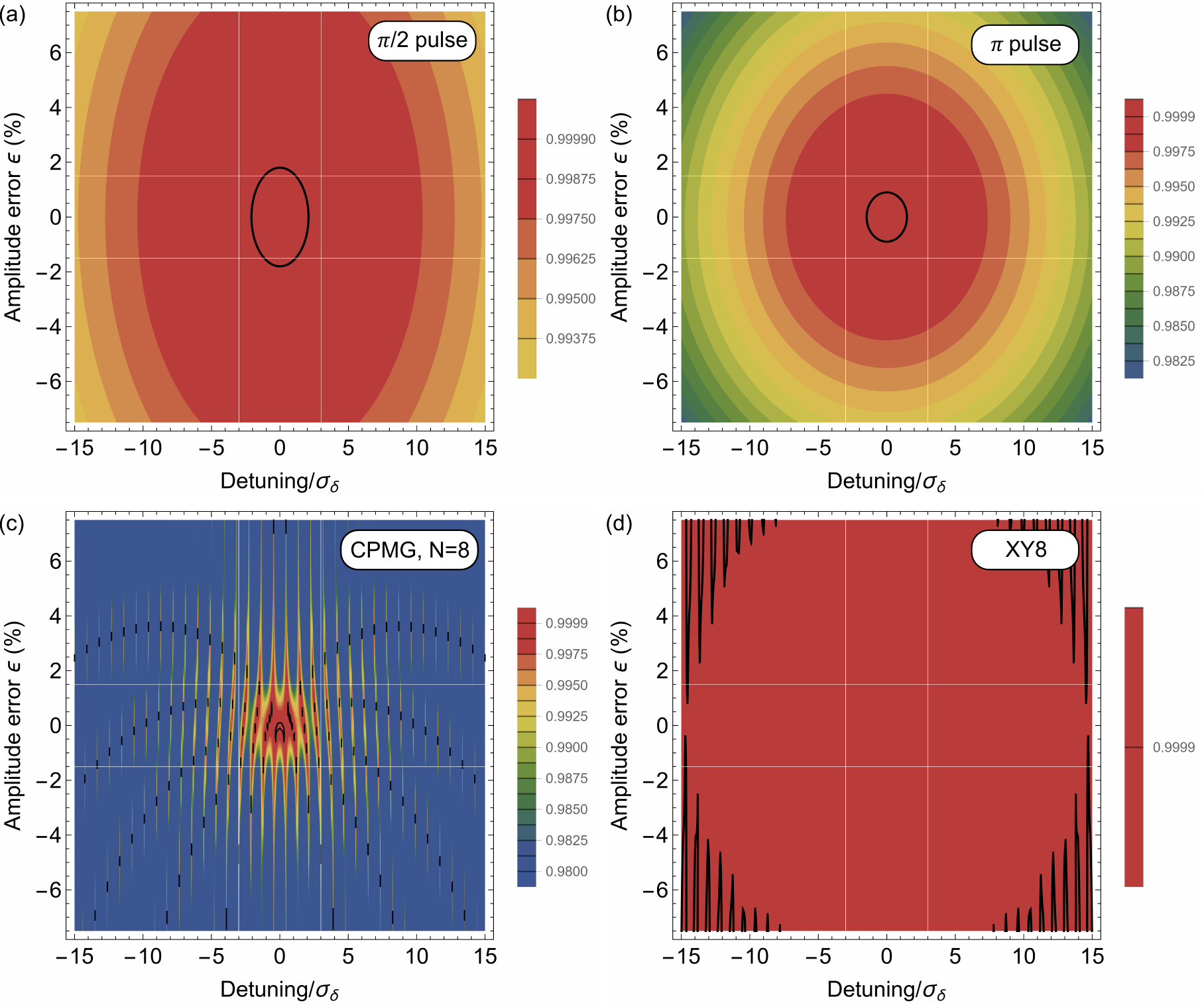}  	    
	\caption{
		Simulation of fidelity of the radio-frequency pulses on the RF1 transition vs. amplitude and detuning errors for (a) a $\pi/2$, (b) a $\pi$ pulse, (c) the CPMG sequence, repeated to include a total of $N=8$ pulses, (d) the XY8 sequence. The time separation between the centers of the $\pi$ pulses of the CPMG and XY8 sequence is 10\,ms, corresponding to the optimum time separation from the simulation in Fig. 3 in the main text. The black lines indicate the typical quantum information threshold of $F(t)=0.9999$. The white lines indicate the range of $\pm 3$ times the standard deviations of the amplitude ($\sigma_{\epsilon}=0.005=0.5\%$) and frequency detuning ($\sigma_{\delta }\approx \sqrt{2}/T_{2,\text{RF1}}^{\ast}\approx (2\pi)112.5$\,Hz), used in the simulations. The target Rabi frequency in the simulation is  $\Omega(t)=(2\pi)11.73\,$ kHz. It is evident that XY8 compensates the expected errors very well. All simulations assume that the electron spin is initialized and remains in state $\ket{\downarrow_e}$. 
	}
	\label{Fig:Fidelity}
\end{figure*}

Figure \ref{Fig:Fidelity} shows a simulation of the fidelity of each of the pulses and sequences vs. amplitude and detuning errors. The results show that the fidelities of the $\pi/2$ and $\pi$ pulses are quite high (Fig. \ref{Fig:Fidelity}a,b). Specifically, the respective fidelities for detuning and amplitude errors at three standard deviations $\delta=3\sigma_{\delta}$ and $\epsilon=3\sigma_{\epsilon}$ are $F_{\pi/2}\approx0.9997$ and $F_{\pi}\approx0.9993$. However, the errors accumulate for CPMG, e.g., when it is repeated and consists of eight $\pi$ pulses (Fig. \ref{Fig:Fidelity}c), with the corresponding fidelity at three times the standard deviations for eight $\pi$ pulses with a $10$ ms pulse separation equal to $F_{\text{CPMG}}\approx 0.96$. On the contrary, the errors are compensated very well for XY8 for the same number of pulses (Fig. \ref{Fig:Fidelity}d), with fidelity practically not differing from 1, demonstrating its robustness. We note that all simulations above assume that the the electron spin is initialized and remains in state $\ket{\downarrow_e}$, i.e., these are the fidelitities of the pulses on the RF1 transition.

We obtain very similar fidelities when the electron spin is initialized and remains in state $\ket{\uparrow_e}$, i.e., for radio-frequency pulses on the RF2 transition. These are $F_{\pi/2}\approx 0.9998$ and $F_{\pi}\approx0.9994$ where the effect of the slightly lower Rabi frequency of $\Omega_{\text{RF2}}=(2\pi)(7.73\pm0.05)\,\text{kHz}$ is compensated by the longer $T_{2,\text{RF2}}^*=(3.35\pm0.21)$\,ms on the RF2 transition qubit. We note that the eigenstates of the RF2 qubit are superpositions of the bare nuclear spin eigenstates due to the strong $A_{zx}$ coupling, which is a particularity of the specific $^{13}$C nuclear spin in the experiment. 
Finally, driving of the nuclear spin, which is independent of the electron spin state can in principle be achieved by simultaneous driving of the RF1 and RF2 transitions, similarly to experiments with NV centers \cite{barry2023sensitiveacdcmagnetometry} or as recently demonstrated for the SnV \cite{beukers2024control}.

\subsection{Heating estimation}\label{subsec:heating_estimation}
\begin{figure}
	\centering
	\includegraphics[width=8.6 cm]{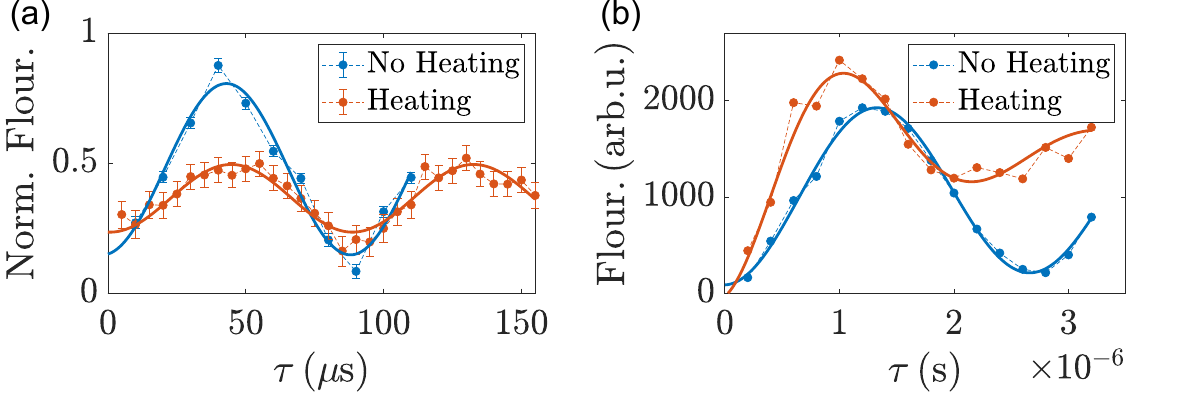}
	\caption{(a) Heating estimation using two nuclear Rabi experiments performed on the same day. The contrast in the nuclear Rabi measurements diminishes when substantial heating is present. For the measurement showing no noticeable heating we deduce a possible pulse spacing of $\widetilde{\tau}=24$\,ms feasible for dynamical decoupling sequences. 
		This value is taken for the estimation of an realistic achievable \nuc memory time.
		(b) The corresponding Rabi normalization shows a detuned driving for the case of significant heating.}
	\label{fig:supp_heating_estimation_mu}
\end{figure}
In our system, we attribute the primary source of heating to ohmic losses caused by impedance mismatches in the MW/RF delivery, particularly at self-soldered connection points near the sample.
Although the wire has minimal contact with the diamond, it is positioned to touch the surface to place the microwave field as close as possible to the color center, which could still transfer some heat to the sample. 
However, we believe the heat load is mainly transferred via the shared thermal ground at the mixing chamber (MXC) flange, where both the sample (via the cold finger) and the MW/RF supply (via connectors) are connected. 
Therefore, in our heating estimations, we focus on the heat load generated by the pulse sequences, specifically the number of RF pulses over a given period. 
\\
For the possible pulse spacing between $\pi$ RF pulses we conservatively estimate $\widetilde{\tau}=2\tau=24$\,ms or longer.
As the temperature sensor states only a low and delayed rise of temperature for our applied sequences, we choose an alternative approach to estimate the heat load in our system. 
For this we compare in Fig.~\ref{fig:supp_heating_estimation_mu} two nuclear Rabi experiments (pulse sequence in Fig.~2\,(c) in the main text) and their corresponding electron Rabi normalization curves.
In order to be able to compare these results in terms of heating with each other, the measurements were taken on the same day with the same parameters, only differing in duty cycles.
We calculate the duty cycle as the ratio of the number of $\pi$ RF pulses to the total duration of the sequence.
For the number of $\pi$ RF pulses we divide the overall duration of applied RF pulses with the $42.5\,\text{\textmu s}$ length of the $\pi$ RF pulse. 
For the duration of the sequence we take the substantial elements, i.e.~waiting and laser elements, into account.
We estimate for the measurement showing no significance of heating a duty cycle of 4.17\,\%.
In the presence of significant heating, we loose contrast in the nuclear Rabi curve (red fit in Fig.~\ref{fig:supp_heating_estimation_mu}\,(a)).
Moreover, the corresponding electron Rabi normalization (Fig.~\ref{fig:supp_heating_estimation_mu}\,(b)) shows detuned driving with a stronger decay. 
We estimate for this measurement a duty cycle of 5.38\%. 
The duty cycles correspond to a (not) possible pulse spacing $\widetilde{\tau}$ for the memory estimation of approximately ($1\frac{\pi_{\text{RF}}\text{-pulse}}{18.6\,\text{ms}}$) $1\frac{\pi_{\text{RF}}\text{-pulse}}{24.0\,\text{ms}}$. 
The corresponding measured temperatures are 88.5\,mK and 93.8\,mK, respectively, compared to a base temperature of approximately 77\,mK taken by a nearby temperature sensor for a RF input power of 5\,W feeded into the cryostat.

\end{document}